\documentstyle [12pt]{article}
\begin{document}
\newcommand{\vm}{\vspace{0.2cm}}
\newcommand{\vl}{\vspace{0.4cm}}

\title{ p-Adic  description of Higgs mechanism II:General
Theory} \author{Matti Pitk\"anen\\ Torkkelinkatu 21 B 39,00530, Helsinki,
 FINLAND}
\date{6.10. 1994} \maketitle

\newpage

\begin{center}
Abstract
\end{center}

\vl

This paper is second one in the series devoted to the calculation of
particle masses in the framework of p-adic conformal field theory limit of
Topological GeometroDynamics.  The concept of topological condensate
generalizes the concept of 3-space  in radical manner.  The various
hierarchically ordered  levels of  the condensate obey effective p-adic
topology and fractal considerations motivate the hypothesis  that
physically interesting values of $p$ correspond to primes near prime
powers of two, in particular Mersenne primes. This  hypothesis relates
succesfully the fundamental elementary particle mass  scales to Planck
mass scale  and what remains is  to predict particle
  mass ratios correctly.   The fundamental description of Higgs mechanism
is in terms of p-adic thermodynamics for the Virasoro generator $L^0$
(mass squared).   The more phenomenological description is based on the
breaking of super conformal symmetry.  In thermal approach   massivation
follows from the small thermal  mixing massless ground state with Planck
mass excitations.   A purely p-adic feature is the quantization of
temperature at low temperature limit, which gives  earlier length scale
hypothesis as a   prediction.   The generalization of $M^4_+\times CP_2$
spinor fields to Kac moody spinors allowing representation of Super
Virasoro algebra and describing both bosons (N-S sector) and fermions
(Ramond sector) is carried out and involves crucially the special features
of $M^4_+\times CP_2$ geometry.  In TGD elementary particles (partons)
correspond to boundary components of particle like 3-surface and
elementary particle vacuum functionals defined in  modular degrees of
freedom of  boundary component are generalized to p-adic context.  The
dominating contribution to elementary
 fermion  and boson masses  comes from boundary degrees of freedom and the
general form of the contribution can be derived by a few line argument.
The calculation of elementary fermion and boson masses is performed in the
third paper of the series.

\newpage

\tableofcontents

\newpage

\section{Introduction}

This is a second paper in the series devoted to the p-adic description of
Higgs
 mechanism
 in TGD \cite{TGD,padTGD} (for p-adic numbers see for instance
\cite{padrev}).   Concerning the general background reader is suggested to
read the introduction of the first
 paper, where general formulation of p-adic conformal field theory limit
was proposed and predictions  were described generally.

\vm

The contents of this paper are following. \\
a)  The physical picture behind p-adic description of  Higgs mechanism is
 described.  The basic elements of description  are  the concept of
topological condensate,  the hypothesis that favoured p-adic primes $p$
correspond to primes near prime powers of $2$,   Super Virasoro
invariance,  thermodynamical and conformal symmetry breaking descriptions
of Higgs mechanism and  the hypothesis that boundary components of
particle like 3-surface are carriers of elementary particle quantum
numbers. For mass  calculations only  the thermal description is
needed.    \\ b)   Super Virasoro symmetry is one of the cornerstones of
the approach and the
 construction of called Kac Moody spinors as generalization of
$M^4_+\times CP_2$ spinors is  carried out.  Kac Moody spinors give rise
to Super Virasoro representations and Ramond/N-S type  representations
describe fermions/bosons. \\ c) Generalization of  massive Dirac operator
is constructed for fermions and bosons and purely p-adic elimination
mechanism for tachyons is found. \\ d) Description of Higgs mechanism as
breaking of conformal symmetry is considered at quantitive level. Some yet
poorly understood problems  (What determines primary condensation level,
What happens in secondary condensation)  are
 considered and the  reader primary interested  in mass calculations can
skip over these sections.  \\
e)  p-Adic counterparts for the so called elementary particle vacuum
functionals, which depend on moduli of boundary component of 3-surface
are constructed.   The construction  involves identification of  p-adic
version of moduli space as a  discrete space and p-adic generalization of
theta functions.  The contribution of boundary component to the mass
squared of particle is calculated.

\vm

 The  calculation of masses of leptons, quarks  and gauge bosons is
carried out in the third paper of the series. It turns out that the
predictions for lepton and gauge bosons masses agree surprisingly well
with known experimental masses.  Errors are below one per cent except for
$Z^0$ boson for which mass is $10$ per cent too large. The reason is too
large value $sin^2(\theta_W)=3/8$ for Weinberg angle:  in the third paper
it is shown that inclusion of Coulombic corrections and topological mixing
effects of leptons leads to a correct prediction for gauge boson
masses.    One can safely conclude that the results of the calculation
verify the essential correctness of  both TGD and its p-adic conformal
field theory limit at quantitative level.

\section{p-Adic description of Higgs mechanism}

The analysis of the weak points of the previous attempts to describe Higgs
 mechanism
 p-adically  \cite{padTGD} suggests what the basic assumptions of the new
 'quark level'  description might be.

\subsection{Super Virasoro invariance}

Super Virasoro invariance is the corner stone of the whole
 approach and follows from the criticality of TGD:eish Universe at
 quantum level ( the concept is described in  the books
\cite{TGD,padTGD}).   In order
 to realize Super Virasoro invariance one must solve several purely
mathematical problems.  One  should construct the representations of Super
Virasoro in $M^4$, $u(1)$ (K\"ahler charge),$ so(4)$  and color degrees of
freedom. Since the spinors of $H=M^4_+\times CP_2 $ play fundamental role
in
 Quantum TGD it seems sensible to try to generalized the concept of
H-spinor to some kind of Kac Moody spinor, which allows the description of
both bosons and fermions as N-S type and Ramond representations
respectively. There are several  problems to be solved. \\ a) The
generalization of $CP_2$ spinors is obviously based on the use of $so(4)$
Super Virasoro representations. The earlier simple description is however
not correct but one must use tensor product of two $so(4)$  Super Virasoro
representations of Ramond type in order to obtain generalization of $CP_2$
spinors. The tensor product of two  $N-S$ type representation  describes
bosons. \\ b) The treatment of $M^4$ degrees of freedom is very difficult
 due to the anomalies: in string models these anomalies lead to critical
dimensions $D=10$ and $D=26$.  In TGD context the solution of the problem
reduces to the observation that $so(4)$ Super Virasoro representations can
be used in the construction of $M^4$ Kac Moody spinors, too.  The
mathematical structures in $M^4$ and $CP_2$ degrees of freedom are
practically identical. It is essential that the product of two
representations is used. The trick is possible due to the fact that
$so(3,1)$ is complexification of $so(4)$ so that dimensions are again
important: now $H=M^4_+\times CP_2$ is the unique alternative. At quark
level all these Super Virasoro representations have $c=h=0$. \\ c)  The
inclusion of $u(1)$ degrees of freedom is simple: one just adds $u(1)$ type
Super Virasoro representation as a tensor factor to the generalized
$H$-spinor. For this representation one has necessarily $c=3/2$. The value
of the
 vacuum weight $h$ gets contribution $h_K=Q_K^2/2$ from the K\"ahler
charge of the particle.
 \\ d) The problem of associating Super Virasoro representation to $su(3)$
degrees of freedom is highly nontrivial. Miraculously it turns out that
$su(3)$ is the only Lie-group allowing $c=h=0$ (color invariance is not
broken), which allows the existence of this kind of representation.
Essential role is played by the symmetric structure constants $d_{ABC}$,
which allow the definition of anticommutator structure in triplet
representation of $su(3)$.

\subsection{Thermal description of Higgs mechanism}

The basic idea is that massless particles become massive via thermal
mixing of the massless ground state  with excitations having Planck mass.
The excitations in question are associated with boundary
 components identified as partons and are labeled by electroweak and color
quantum numbers.   Mass
 squared contains also interior contribution but this need not allow
thermal description. For instance, states of higher spin
 on hadronic Regge trajectories correspond to to interior excitations so
that thermal  description predicts only the ground state masses of
hadrons.
 Thermal description  of the mixing applies  in principle for all values
of prime $p$.
   Thermalization takes place  in Virasoro degrees of freedom  $L^0$
being analogous to energy. It turns out that thermalization cannot be
significant effect for spin, electroweak and color quantum numbers.
Virasoro and isospin/spin degrees of freedom separate nicely for coset
representation used to construct Kac-Moody spinors.
 There are obvious  constraints on  quantum numbers of,  say,  hadron. One
can require  that the over all ground state spin of the hadron is fixed,
that hadron is color singlet, etc.

\vm

The quantization of temperature ($T=1/k$) makes scenario
 predictive and for large values of $p$ predicts correct mass scale
$O(p^k)$ ($k=1$ seems to be the most interesting value of $k$).   This
prediction
 in turn implies that $m$  differs from $m=2$ by order $ O(p^{k+1})$
contribution. For Ramond (N-S) representations  $P,Q,Q_K$ differ by
order   $O(p^k)$ ($(p^{k+1}$ ) contribution from their values for massless
representations.    For $m=2$ representations  with $h=0$ mass squared  is
in extremely good approximation given by the contribution of $n=1$ excited
states to thermal average and is proportional to some degenerary factor
multiplying $p^{n/T}$ factor.

\begin{eqnarray}
\langle m^2\rangle &\simeq & k \sum_{n=1,2}A(n) p^{n/T}\nonumber\\
1/T&=&1,2,....\nonumber\\
A(1)&=& \frac{D(1)}{D(0)}\nonumber\\
A(2)&=& \frac{(2D(2)-\frac{D^2(1)}{D(0)})}{D(0)}
\end{eqnarray}

 \noindent Here $D(i)$ denotes the degeneracy of $M^2=i$ state.  $k$ is
proportionality constant, which turns out to have value
 $k=\frac{3}{2}$.
 For $T=1$ mass squared scale is $O(1/p)$ and for $T=1/2$  mass squared
scale is $O(1/p^2)$.    It turns out that $T=1$ is the value of
temperature for fermions  and $T=1/2$ for gauge bosons.  The powers $p^n$,
$n>2$ give completely negligible contribution to mass squared so that
computational
 inaccuracies are under very tight control.

\vm

There is very delicate and fundamental p-adic effect involved.
 If the coefficient $kA(1)=kD(1)/D(0)$ of order $O(p)$ contribution is not
 integer the mass of the state is practically Planck mass. Therefore only
those states for which the condition

\begin{eqnarray}
\frac{kD(1)}{ D(0)} \  \ integer
\end{eqnarray}

\noindent  holds true are light. The degeneracies of massless states turn
out to be larger than one due to the fact that ground state is p-adic
tachyon with $M^2(ground)$ being negative integer or half odd integer. The
same mechanism implies also that second order contribution to mass squared
cannot be neglected unless $kA(2)$ is integer.  This contribution is
present in massive N-S and Ramond representations assumed to provide
phenomenological description of Higgs mechanism. Clearly  the degeneacy of
massless states is analogous to vacuum degeneracy related to the direction
of Higgs vacuum expectation value.  Of course, Higgs field before symmetry
breaking is also tachyon.

  \vm

A further delicate p-adic effect is the fact that p-adic tachyons $M^2<0$
 do in fact correspond to real p-adic masses due to the delicacies of
p-adic square root under certain conditions for $p$: Mersenne primes
belong to those p:s. What happens is that would be tachyons are actually
Planck mass particles.

\vm

 At which instances thermal description makes sense?  Certainly
 it looks natural to assume that the mass of the particle at the  primary
condensation level is determined thermally.   Since in TGD elementary
particle quantum numbers reside on boundary components,    boundary cm
degrees of  freedom, which are carriers of standard model  quantum
numbers,  should be thermalized.     A
 suitable generalization of the  thermal description to the conformally
invariant degrees of freedom of
 the boundary component (characterized by Teichmuller parameters) is  also
expected to make  sense: it turns out that modular degrees of freedom
give contribution to the vacuum weight of the
  Super Virasoro representation associated with boundary cm degrees of
freedom.  As already
 noticed, interior contribution  to mass squared should be identifiable
as   counterpart for the
 known contributions  such as   Coulombic  and color Coulombic binding
energy, spin-spin interaction
 for quarks, etc...  to particle mass and need not      have any thermal
description.

\vm

Thermal description for the secondary condensation of massive particles
(Super Virasoro representations with vacuum weight $h>0$) is not useful.
For  purely number  theoretic reasons one must assume $T=
\frac{1}{8m(m+2)N}$ , where  $N $
 and $m$ are integers and  $m=2+O(p)$  characterizes  the Super Virasoro
representation.  The extreme smallness of the temperature  implies that
the contribution of the  excited states to mass squared is proportional
 to $p^{m(m+2)N}$ and extremely small for large values of $p$.   This
means that   thermal effects
 cannot explain the   condensation energy
 assumed to be  associated with secondary condensation in \cite{padTGD}.
 Probably condensation energy corresponds to the  change in Coulomb self
energy, etc.   associated  with the interior of the particle like
3-surface.   Neither does the  use of thermal description
 resolve the problem of what happens to the values of p-adic quantum
numbers $P,Q;m,Q_K$ associated with massive Super Virasoro representation
in the condensation
 of level $p_1$ to level $p_2>p_1$.

\subsection{Higgs mechanism as breaking of Super Conformal Invariance}

The description  of Higgs mechanism as breaking of conformal symmetry
 parametrizes the
 predictions of fundamental thermal description at primary condensation
level.   The new element is the inclusion of electroweak symmetry breaking
by decomposing
 the  representations to the representations of $u(1)_V \times u(1)$.
Equivalent  description is in terms of central charge $c$  and vacuum
weight $h$,  which are linear combinations  of $I$ and  $I_3$.   Correct
order of magnitude for mass scales result if the deviations for
 $so(4)$ parameters are of order $O(p)$ and satisfy certain additional
constraints.   Also the order
 $O(p^2)$ contribution turns out to be important. If exact conservation of
electromagnetic charge
 is assumed $u(1)$ charge is not free variable.

\vm

 Thermal description,   when applied  at the  primary condensate level
should predict elementary particle and hadron ground state  masses and
it should be possible to deduce
 the values of  the parameters $P,Q,m$ associated with the  corresponding
symmetry broken Super Virasoro representations. The basic formula for mass
squared

\begin{eqnarray}
\langle M^2(h=0)\rangle_{thermal} &=& M^2(h(P,Q,m)\neq 0)
\end{eqnarray}

\noindent  indeed gives a constraint on the parameters $P,Q,m$ of the
symmetry
 broken $h\neq 0$ representation.    The interactions associated with the
interior of
 the particle like 3-surface   (Coulomb self interaction,etc..)  give also
 contributions to the mass squared. The simplest possibility is that
interior contributions  affect  only the vacuum weight $h$ and therefore
the parameters $P,Q,m$  of the symmetry broken Super Virasoro
 representation associated with particle.   For hadrons interior
contributions are large and  responsible for the masses for the higher
spin states on hadronic Regge trajectories and
  parameters $P,Q,m$ of the symmetry broken  Super Virasoro representation
associated with hadron  (tensor product of quark representations)  should
contain additional contribution, which  depends on  hadronic spin and is in
lowest order linear in hadronic spin so that  standard mass formula for
Regge  trajectories is obtained.   Contrary to the
 original beliefs \cite{padTGD},  hadronic string tension is  not
predictable    from  the thermal approach  but necessitates a model for
 the interior degrees of freedom (perhaps TGD:eish counterpart \cite{TGD}
of the old fashioned string
 model).

\subsection{Electroweak mass splitting in boundary degrees of freedom}

The description of electroweak splitting in bosonic case requires
 no special assumptions. In fermionic case the mass squared operator must
contain a part with depends on the isospin  of the fermion. The kind of
term is indeed allowed since Super Virasoro representation can always be
decomposed to Super Virasoro representations associated with the vectorial
$u(1)$ subgroup of $SO(4)$ so that the addition of isospin dependent term
just shifts the vacuum weights $h$ of different Virasoro representations
in isospin dependent manner.  The vacuum weight can be expressed in the
form

\begin{eqnarray}
h&=&h_0+dI_3
\end{eqnarray}

\noindent $d$ must be integer: otherwise Planck mass results. $d=1$  turns
out to be the correct value of $d$ and that it turns out that  vacuum
weights are are given by

\begin{eqnarray}
h(\nu_L)=h(U)&=&-\frac{5}{2}\nonumber\\
h(L)=h(D)&=& -\frac{3}{2}
\end{eqnarray}

\noindent  Since $M^2=n$ states are created by operators of weight
$\Delta=n+h$ the thermal averages are different for charged lepton and
neutrino and thermal mass splitting results. For bosons isospin dependent
 term in mass squared operator is excluded by CP invariance.

\subsection{Higgs mechanism without Higgs}

In the standard description of Higgs mechanism massive gauge bosons
receive their longitudinal polarization from Higgs field.
  In the description of Higgs mechanism as a breaking of super conformal
symmetry there is no need for Higgs field.   Kac Moody spinors ccontain as
tensor product factor  discrete series  representations of Super Virasoro
in $so(4)$ degrees of freedom.  The corresponding vacuum weight
$h(P,Q,m=2+\Delta m)$ differs from its value at massless limit  $\Delta
m\rightarrow 0 $ so that  the norm of  the longitudinal polarization is
proportional to $p^2\propto \Delta m $ and nonvanishing.

\vm

 In the thermal description of Higgs mechanism using $h=0$
representations one can define the norm of the  massless longitudinal
polarization  states as the  limiting value of the norm for $\Delta m\neq
0$ representations (this limit makes sense only p-adically!).   One cannot
exclude the possibility that  in actual physical situation secondary
topological condensation implies a small deviation of $m$ from $m=2$ so
that limiting procedure is only a convenient mathematical idealization of
the actual situation.  The ordinary description of Higgs mechanism indeed
suggests that the $h=0$ representation is unstable against  generation of
small value $\Delta m$.

\subsection{ Universal string tension}

 String tension $M_0^2$ does not depend on $p$ and is apart from
 some numerical constant equal to $1/G$ as in string models.   String
tension itself  is prediction of Quantum TGD and involves  the theory in
Planck length scale.   It turns out that $T$ is of same order as the
string tension associated with cosmic strings in GUTs. p-Adic
thermodynamics at low temperature limit predicts the mass scale at
condensate level $p$ to be given by $M^2(p)=M_0^2/p$.    The description
simplifies drastically the description of elementary particles.  There are
no problems with light excitations  (as in previous scenario) since they
are totally absent. Primary condensation  levels can be assumed  to
correspond  primes, which define hadronic and leptonic mass
 scales that is Mersenne primes and primes near prime powers of $2$.

\vm

As already noticed the ordinary hadronic string tension emerges in the
conformal symmetry breaking description naturally via the dependence of
the parameters $P,Q,m$ on the  spin of the hadron. Since  interactions
related with the interior of the hadron are involved   thermodynamic
description cannot predict its value.

\subsection{ Boundaries as carriers of elementary particle quantum numbers}

In TGD fermion families correspond to different boundary topologies
 characterized by the genus $g=0,1,2,..$ of the boundary component.
  The natural expectation is that boundaries give the dominating
contribution to the
 thermal expectation of the mass squared operator. There are several
questions to be answered?   Does the entire contribution to mass come from
boundaries?  Can one separate boundary  contributions to cm contribution
and contribution associated with conformally invariant degrees of freedom
(modular degrees of freedom characterized by Teichmuller parameters)?
 Are cm and modular  contributions independent and additive?  Can one
estimate modular contribution using thermodynamic approach or should one
estimate it as quantum mechanical expectation value? Hard work with
various alternatives has taught that answer is very simple.  Mass squared
for elementary particles is  in good approximation sum of two
contributions

\begin{eqnarray}
M^2&=&M^2(cm)+M^2(mod)
\end{eqnarray}

\noindent The first contribution correspond to either cm of boundary
 component and depends on 'standard' quantum numbers.  The second
contribution  is
 the contribution of modular degrees of freedom and is same for all
fermions  and depends only on the genus of boundary component.

\vm

  The most pleasant surprise of calculations was that $M^2(mod)$ can be
deduced apart from overall integer valued proportionality constant from
the general form of elementary particle vacuum functionals  constructed in
\cite{TGD}. Together with the knowledge of cm contribution and Mersenne
prime hypothesis one can predict lepton and gauge bosons masses with
accuracy better than one per cent.

\vm

The actual calculation of $M^2(mod)$  necessitates the introduction
 of p-adic generalization for the space of moduli (Teichmuller parameters)
as well as construction of modular invariant  partition functions plus
exact definition of the  mass squared operator.  It turns out that
surprisingly simple p-adication of moduli space as discrete space exists
and that   that elementary particle vacuum functionals as such provide the
solution to the problem!  In particular, the earlier arguments explaining
the absence of $g>2$ generations remain valid as such.  Amusingly enough,
at this stage it is not  clear whether one should regard modular
contribution as quantum mechanical  of thermal expectation value of the
mass squared operator.

\vm

The harder part of the calculation is the evaluation of the cm
contribution $M^2(cm)$.  Fortunately, this contribution is same for all
charged leptons/neutrinos. The contribution  of quarks and leptons are
also almost
 identical.
  It turns out that there are many delicacies related to the concept of
 Kac-Moody spinor: in particular the definition of the inner product
requires special care. The degeneracy of massless states turns out to be
generic phenomenon very analogous to the vacuum
 degeneracy of Higgs field and light fermions  are miracles in the sense
that the ratio $D(1)/D(0)$ is integer for them.

\vm

What is the size of the boundary components associated with elementary
 particles?  The  belief has been that boundary components have size of
order Planck length.  The fact that elementary particle  vacuum
functionals depend on the conformal equivalence class of the particle only
suggests that there is no preferred size.  The success of p-dic  mass
formula in turn suggests that boundary components in fact correspond to
the outer boundary of the small approximately flat piece of Minkowski
space associated with the elementary particle and that the average size of
surface is of order p-adic length scale and hierarchy of boundary size
scales results.  Hadron physics considerations gives additional support to
this belief.

\vm

The mixing of boundary topologies is also possible.  It turns out that
Cabibbo-Kobayashi-Maskawa mixing can be understood  essentially as
difference of mixings of $g=0,1,2$ topologies  for $U$ and $D$ type
quarks.

\subsection{ Open problems}

There poorly undestood aspects  of TGD:eish description of  Higgs
mechanism can
 be condensed on few basic questions. What determines the primary
condensation level?  Why primes
 near prime powers of two are favoured?   What happens in secondary
condensation?

\subsubsection{What happens to symmetry breaking parameters in secondary
topological condensation?}

  The problem what happens in the secondary condensation of condensate
level  $p_1$  to condensate level $p_2$ belongs outside the context of
p-adic conformal  field theory and one can make only guesses at this
stage.   It makes no sense to  relate to each other directly the values of
various p-adic quantum numbers: only their real  counterparts are
comparable.  What looks physically plausible  is that mass decrease is
necessary for  condensation to occur.  Concerning the details of
condensation one can make  only guesses at this stage. An attractive
possibility is that some kind of variational principle  is involved.   \\
a) An untested hypothesis   of \cite{padTGD} is that real mass decreases in
 condensation and the decrease is as small as possible.  This principle is
essentially equivalent with the p-adic version of  the
statement that entropy gradient $dS/dt$ is as small as possible in time
evolution  since mass squared expectation value is
for  discrete p-adic temperature essentially equal to  entropy.   This
requirement in principle  determines the renormalization of the quantities
$P,Q,m$  characterizing the conformal  breaking.  The problem is that without
 any
constraints on the change of parameters $P,Q,m$ the mass change
 can be made practically zero. Of course, one can consider the addition of
various physical
 constraints.  \\ b)  One possibility is that the real counterparts of
$\Delta P, \Delta Q  $
 and $\Delta m $ are 'small' but the problem is how small. An other
extreme would be that the real counterparts $P_R,Q_R, m_R)$  of
 $P,Q,m$ remain invariant in condensation:  this could be interpreted as
conservation law in some sense and might well minimize the mass change.
This implies that mass change  results completely from the fact that   for
an algebraic expression $f(P,Q,m)$

\begin{eqnarray}
f_R(P,Q,m)&\neq & f(P_R,Q_R,m_R)
\end{eqnarray}

\noindent This alternative is formally  appealing since mass change
 is guaranteed to be small and  the mass change is analogous to quantum
corrections resulting from noncommutatitivy of quantum counterparts of
classical observables.  Furthermore,  the mass change might be also
positive: in this case the condensation obviously is not stable. Could it
be that favoured
 condensation levels are just those for which the condensation energy is
negative?

\vm

There is  a counterargument against both scenarios. In \cite{padTGD} the
 concept of condensation energy was introduced.  The identification of the
condensation  energy as the change induced by the condensation to the
various interior contributions (Coulomb  self energy, etc.)  to the vacuum
weight $h(P,Q.m)$ looks natural.     This suggests that  condensation
affects the values of $P_R,Q_R,m_R$ on the new condensation level so that
scenario b) could not be correct as such.
 In the last paper of the series it will be found that the  real
counterpart of the  p-adic  gauge coupling $g^2$ can  be defined via
canonical
 identification $g^2_R=(g^2p)_R$  assuming $g^2$ is rational number. The
value of the real  counterpart is very sensitive to the value of the prime
$p$.  If  the effective values of gauge couplings are indeed sensitive to
the value $p$ then condensation can affect the mass of  the particle in
decisive manner and   lead to a  dynamical selection of condensation
levels and explain why Mersenne primes and primes near prime powers of two
are favoured physically.

\subsubsection{  What determines the primary condensation level?}

  The basic assumption is that there exists some critical prime  $p_{cr}$ so
that for $p<p_{cr}$ it does  not  make sense to speak of the condensation of
massless particle but secondary condensation of massive particle is in question
and the change of particle mass in condensation can be regarded as
renormalization of mass.   What   principle determines  the value of $p_{cr}$?
What stabilizes the primary condensation level $p_{cr}$?
  Why some condensation levels seem to be in
special position physically (the considerations of \cite{padTGD} suggest
strongly that primes near prime powers of $2$ are such primes)?   One can
imagine  several alternative answers to these questions.

\vm

    p-Adic thermodynamics might provide  the
principle selecting the favoured  primary condensation levels. The maximation
 for
the real counterpart of p-adic entropy is essentially equivalent with
the minimization
of mass squared as a function of $p$ and since second order term in mass
 squared
depends very sensitively on $p$,  local mimina of mass squared are expected to
occur and principle  leads to favoured primes. An open question is
whether the principle is strong enough to fix primary condensation level
sufficiently uniquely. If one requires absolute mass
minimum then there is no upper bound for $p_{cr}$ since mass behaves as
$M^2\propto \frac{1}{p}$.  Therefore the correct question to ask is  what
stabilizes certain primary condensation level  so that
 primary  condensation levels  with $p<p_{cr}$ are not possible. Of course,
one cannot exclude the possibility of several values of $p_{cr}$
and in the  fifth paper of the series the   possibility of  scaled up copy
 of
ordinary hadron physics ($M_{107}$)   at  $M_{89}$ condensate level is
considered.

\vm

A possible stabilization mechanism  is  following.  \\
a) Massless elementary particles can in principle condense  on  several
 primary condensate levels $p\ge p_{cr}$.  The condensation  on  levels with
$p>p_{cr}$  is   not however stable  since the mass is always  larger than at
$p=p_{cr}$ level.  What stabilizes $p_{cr}$ is that real mass squared is
smaller
for   the  primary condensation on $p_{cr}$ followed by  secondary
condensation  on some level $p_0<p_{cr}$ near $p_{cr}$  than for direct
 primary
condensation  on $p_0$.
 Same principle applies to further secondary condensations, too  and
might explain   why  certain primes  are in preferred position physically.
The
mass calculations for hadrons give support for this mechanism:  both the
primary
condensation level $p_{cr}$ for $u,d,s$ quarks and condensation level of
 hadron
(secondary condensation level for quarks)  correspond to primes near
 $2^{107}$.
Note that this principle is in accordance with local entropy maximation
principle.
 \\ b) A more quantitive description is obtained by adopting either
the minimization of $\Delta M^2$ in secondary condensation or the hypothesis
 that $P_R,Q_R,m_R$ remain invariant in condensation.  Later the latter
possibility will be studied in more detail.

\vm

Consider next the question why primes near prime powers of two are favoured
physically. \\
a) Period doubling analogy suggests one partial explanation.  The underlying
2-adic fractality suggests a second  partial explanation: scales
 $2^nL_0$
are fundamental and
are  essentially identical with scales $p^nL_0$  (synchronization) if $p$ is
near  power of $2$.\\
b) p-Adic  entropy maximation principle: the real counterpart of mass squared
 as
a function of $p$ has  local  minima for $p$ near prime power of $2$. This
conjecture is   testable via the study of the  second order
term in mass squared. \\
  c)  The
following argument suggests a  partial answer to the question
 why primes near powers of $2$, especially Mersenne primes are favoured.
The  inverse temperature associated with condensed particle can be any
non-negative integer $n$.  For large values  of $n$ ( small temperature
and small mixing with Planck mass excitations)   also particles of small
$p$ can condense with small mass. This suggests the possibility that the
value of $n$ depends on $p$ so that the value of the mass doesn't depend
very much on $p$.  For primes $p$ near powers of $2$: $p \simeq 2^k$ it is
indeed possible to get essentially identical masses by suitable choice of
$n$: $p^{k}\simeq  2^{nk}$.  For instance electron mass scale would
correspond to 2-adic temperature $T = 1/127$.   The difference between
2-adic masses and p-adic masses is smallest when primes $p$ is as near as
possible to power of $2$ and this  condition selects Mersenne primes as
especially favourable candidates for $p$. \\ d) The real counterparts of the
various coupling constants $g^2_i$  (such as  $e^2,g_{ew}^2,...$) turn out to
correspond to the real counterparts of $g^2_ip$ in canonical correspondence.
If
$g^2_i$ is finite superposition of powers of $2$  and $p$ is Mersenne prime,
the real counterpart is what it should  be:  for other primes this is true
 only
under rather restrictive conditions. This means that the effective values
of gauge couplings and other parameters depend sensitively on condensate
level and this probably gives  dynamical reason for the special role of
Mersenne primes.   It turns out also, that Mersenne primes very probably
correspond to fixed points of color coupling constant evolution: there is
one QCD for each Mersenne prime as was suggested in \cite{padTGD}. The
critical value of $p$  is  analogous to critical temperature since the
real counterpart of p-adic temperature is proportional to $1/ln(p)$ so
that primary condensation is phase transition like phenomenon.

\section{Super Virasoro symmetry }

 The existence of the p-adic super Virasoro representations poses a
stringent
 consistency test for TGD.  \\ a)  In $M^4$ degrees of freedom the
dynamical degrees of freedom involve
 quantized $M^4$ coordinate and Kac Moody version for spin degrees of
freedom.   The fact that
 spacetime dimension $D$ is four implies that $M^4$ coordinates are in fact
redundant dynamical variables for space times surfaces representable as
graphs for a map $M^4\rightarrow CP_2$.  This in turn  suggests the
vanishing of the conformal anomaly  $c$ now proportional to $4-D$, $D$ the
dimension of spacetime surface.  The breaking of Lorentz invariance in
string models for noncritical imbedding space dimension results from the
nonvanishing of conformal anomaly and is expected to be absent for the
simple reason that $M^4$ and space time surface have same dimensions.
The elimination of  vibrational coordinate degrees of freedom leaves only
the Kac Moody version for spin degrees of freedom and the remaining  task
is to construct generalization of $M^4$-spinors.  \\ b)   In $CP_2$
degrees of freedom one has both coordinate and spinorial degrees of
freedom.  Spinorial degrees of freedom correspond to $CP_2$ part of
generalized H-spinor.   Coordinate degrees of freedom  correspond to Super
Virasoro represention for color group.  \\ c)  The existence of spinor
structure in $CP_2$ forces a modification of spinor structure via gauge
coupling to K\"ahler potential.  The counterpart  of  $u(1)$ coupling  is
$u(1)$ Super Virasoro representation appearing as tensor factor of Kac
Moody spinor.

\vm

The basic  guideline in the search for realization of Super Virasoro
algebra is  clearly the need to generalize the concept spinor field  of
$H=M^4_+\times CP_2$  to 'Kac Moody' spinor allowing  realization of Super
conformal symmetry for $so(3,1)$, $so(4)$, $u(1)$ and $su(3)$.
 Also the description of bosons should be possible using  Kac Moody
spinors:
 fermions and bosons would correspond to Ramond and N-S type
representations.

\vm

Ordinary spinors are constructed as tensor products of 2-dimensional
spinors
 and same construction should apply now.  The analog of two-dimensional
spinor is clearly single  $(c,h)=(0,0)$  $so(4)$  Kac Moody representation
 \cite{Goddard,Bible}, which decomposes into sum over tensor products of
unitary  $su(2)_V$ ('V' refers to vectorial)  Kac Moody  and Super
Virasoro representations \cite{Goddard,Bible}  and  for which vacua form
isospin doublet  in Ramond case and isospin triplet and singlet in N-S
case.  Four-fold tensor product of $so(4)$ representation should give rise
to both  $M^4\times CP_2$ Kac Moody spinors.  Tensor product can contain
only Ramond or N-S type representations and this implies  that  only spin
$(J,I)=(1/0,1/0$ vacuum states are possible for N-S sector and
$(J,I)=(1/2,1/2)$ vacuum states in Ramond sector. In particular spin 0
isospin doublet (Higgs doublet) is excluded.

\vm

The  description of $u(1)$ and $su(3)$ quantum numbers is in principle
 straightforward. One must add appropriate $u(1)$ and $su(3)$  Super
Virasoro representations to the Kac Moody version of H- spinor as a tensor
factor.  $u(1)$ representations should be of  type Ramond/N-S   for
fermionic/bosonic Kac Moody spinors.  A  good guess is that color Super
Virasoro is of Ramond type for color triplet vacuum state (quarks)  and
N-S type for color singlet vacuum (leptons). This raises a little
technical problem.  Leptons corresponds to tensor products of Ramond type
and N-S type representations: how can one realize Ramond type Super
Virasoro for this kind of structure.  It turns out that problem can be
solved.

\vm

The existence of required Super Virasoro representations is not at all
obvious.\\ a) The transition from $CP_2$ Kac Moody spinors to $M^4$ Kac
Moody spinors
 need not be trivial although the tangent space groups $so(3,1)$ and
$so(4)$ are related by complexification. The dimensions of $M^4$ and
$CP_2$ are in fact crucial for the concept of  Kac Moody spinor to make
sense.\\ b)   It could happen that  color group does not allow Super
Virasoro representation of required type:  color would be totally
invisible  in the conformal field theory limit of TGD.  In the previous
work only Super Virasoro representations describing color confined states
were considered  (fit of particle masses \cite{padTGD})  or the color
degree of freedom was not treated explicitely (2-adic description of Higgs
mechanism \cite{padTGD}).     The fact that quark level description is
empirically found to work so well suggests that  color Super Virasoro with
physically acceptable properties  indeed exists.  To my personal knowledge
the existence of  this kind of representation has not been documented in
literature and the motivation for the construction to be described came
from TGD.   It came as  a pleasant surprise to find that the very special
properties of $su(3)$  (and  therefore of $CP_2$) allow the construction
of this representation.

\vm

In the following the construction of the required Super Virasoro
 representations is described. \\ a) The most  relevant properties  of
$so(4)$ Super Virasoro representations are
 reviewed. \\ b) Realization of Super Virasoro  invariance for $M^4\times
CP_2 $ Kac Moody  spinors is considered. \\ c) $u(1)$ Super Viraroso is
described.  \\ d) Color Super Virasoro  representation  is constructed:
crucial role is played by the completely symmetric structure constants
$d_{ABC}$ existing for color group only.\\ e) The identification of
elementary particles as generalized spinors is summarized with  comparison
with more standard theories.

\subsection{ Relevant  properties of $so(4)$  Super Virasoro representation}

   $SO(4)$ is quite unique  among other Lie groups in that it admits all
unitary discrete series  representations of Super Virasoro algebra
\cite{Goddard,padTGD}.
 $su(2)_V$ Super Virasoro representations are obtained using
$su(2)_L\times su(2)_R/su(2)_V$ ($V$ refers to vectorial subgroup)
  coset  construction \cite{Goddard} and can be decomposed into a  sum of
tensor products of Super Virasoro and $su(2)_V$ Kac Moody representations
characterized by the isospin (or highest weight $q$) of the vacuum
representation and by central charge $k_V$.   This decomposition is of
central importance since it implies separation of Super Virasoro and
$su(2)_V$ Kac Moody  degrees of freedom and plays central role in the
construction of physical states and mass calculations.

\vm

$su(2)_R$ Kac Moody representation has highest weight $p$ and central
charge  $k_R$. $su(2)_L$ Kac Moody representation has central charge
$k_L=2$ and decomposes corresponds to isospin $1/2$ (Ramond) or direct sum
of isospin $I=0$ and $I=1$ (N-S).  The basic formulas  for the  conformal
central charge $c$ and the allowed vacuum
 weights $h$ of  general discrete series Super Virasoro representation
appearing in tensor product composition read as:

\begin{eqnarray}
c&=& \frac{3}{2}(1-\frac{8}{m(m+2)})\nonumber\\
h&=& \frac{( (m+2)p-mq)^2 -4}{8m(m+2)} +\frac{\epsilon}{16}\nonumber\\
m&=& k_R+2\nonumber\\
\epsilon&=&1(0) \ for \ Ramond (N-S)\nonumber\\
\
\end{eqnarray}

\noindent  Here one has $p=1,..,m-1$ and $q=1,..,m+1$. $p-q$ is odd for
Ramond representation and even for N-S representation. $p$ and $q$ are  the
highest weights for $su(2)_R$ and $su(2)_V$  Kac Moody representations.

\vm

As far as thermal description of Higgs mechanism are considered the most
interesting representations are $(c,h)=(0,0)$ representations with unbroken
conformal symmetry. These should correspond to massless particles.  There
are three representations with $(c,h)=(0,0)$.  They have  $m=2$ and
therefore vanishing central charge  $k_R=0$ for the second $su(2)$ factor
of $SO(4)$.   In Ramond sector the representation decomposes to tensor
product  of Super Virasoro and  $su(2)_V$ Kac Moody representation
having  $I=1/2 $  ($(p,q)=(1,2)$).    In Neveu-Schwartz sector the
representation decomposes into a tensor product of Super Viraroro with
sum of    $I=0$ ($(p,q)=(1,1)$)
 and $I=1$ ($(p,q)= (1,3)$) representations.

\vm

The  identication of Kac Moody spinors as 4-fold tensor product of
 $(c,h)=(0,0)$ $so(4)$   Super Virasoro representations allows only spin
$1/2$  isospin  $1/2$ elementary  fermions and spin $0/1$  isospin $0/1$
elementary bosons  in accordance with experimental facts.
  Higgs doublet would require the tensor product of N-S ($M^4$ and Ramond
$CP_2$ type
 representations and this probably leads to difficulties with the
realization of Super Virasoro symmetry (one should somehow transform the
N-S type generators to Ramond type generators in $CP_2$ type tensor
factors).

\subsection{Super Virasoro invariance in $M^4_+\times CP_2$ spinorial
degrees of freedom}

 There are strong physical  constraints on the realization of Super
Virasoro
 invariance in $M^4$ degrees of freedom. \\ a) $(c,h)=(0,0)$
representation  is required by Lorentz invariance. In the standard
realization of string models  in terms of bosonic and
 fermionic fields one encounters the well known problems.  In present case
however the dimension of spacetime surface is $D=4$ and identical to the
dimension of $M^4$ and since p-adic field theory limit is expected to make
sense at the flat spacetime limit only, $M^4$ coordinate degrees of
freedom can be eliminated almost totally: only  cm degrees of freedom
remain and the generator $L_0$ gets only the $p\cdot p$ contribution from
$M^4$ coordinates. Fermionic degrees of freedom are not lost however and
are necessary for describing the spin degrees of freedom. \\ b)  The
realization should provide some kind of local version of Lorentz group and
respect Lorentz invariance. String theories have demonstrated how
difficult this requirement is to satisfy.\\ c)  The representations should
allow generalization of $H$-spinors to Kac
 Moody context and also the generalization of Dirac equation. The formal
analogies between Ramond and N-S representations suggest that the spinor
 concept and  Dirac equation should also generalize to the N-S case.

\vm

The need to obtain generalization of H-spinors allows clue to the
construction. The crucial observation is that single $so(4)$ coset
representation for
 fermions gives just one $J=1/2$ doublet.  What is needed is two doublets.
The only manner to achieve this is to use tensor product of two $so(4)$
coset representations. Same applies in $so(4)$ degrees of freedom so that
4-fold tensor product of $so(4)$ coset representation is needed in order
to generalize the concept of H-spinors.  Lorentz generators cannot be
constructed in
 terms of single $so(4)$ representation but it turns out that the tensor
product of two representations allows to construct the local version of
Lorentz algebra in terms of commutators of the  gamma matrix like
operators $F^{am}_i, i=1,2$ and the construction respects Super conformal
invariance.

\vm

The result means that there is amazingly close mathematical
 relationship between $M^4$ and $CP_2$  degrees of freedom.  The result is
 unique to the 4-dimensional Minkowski space so that construction provides
one further argument in favour of 8-dimensional imbedding space and its
decomposition into $M^4$ and $CP_2$  factors.

\subsubsection{Consistency of Lorentz and Super conformal symmetries}

The previous construction does not make manifest the consistency of Lorentz
 and super conformal invariance. The consistency can be however seen by
detailed construction  of Lorentz algebra in terms  $so(4)$ coset
representation.  Consider first Ramond sector. The counterparts of Dirac
gamma matrices  can be identified in terms of super generators $F^{a0}_i$,
$i=1,2$  of $so(3,1)$  sector

\begin{eqnarray}
\gamma^0&=&iF^{1,0}_1\ \ \gamma^3=iF^{2,0}_1\nonumber\\
\gamma^{1}&=&iF^{1,0}_2\ \ \gamma^2=F^{2,0}_2
\end{eqnarray}

\noindent The corresponding identification for the counterparts of  $CP_2$
gamma matrices in sectors $i=3,4$ is obviously  possible
 and the only difference is  the absence of imaginary unit.  The
construction generalizes to bosonic case with the difference that now
gamma matrices are $\Delta =1/2$ operators $\gamma^k_{1/2}$ constructible
from $F^{a1/2}_i$ and their conjugates $\gamma^k_{-1/2}$ can  be defined
in terms of $F^{a,-1/2}_i$.

\vm

Sigma matrix representation for Lorentz algebra  for Ramond representation
 can be defined as commutators of gamma matrices in standard manner. In
bosonic case one must define sigma matrices as commutators
$\gamma^{k}_{1/2}$ and $\gamma^l_{-1/2}$ in fact localization is in both
cases possible matrices $\Sigma^{kl}_n= \sum_m
Comm(\gamma^k_{n-m},\gamma^l_m)$. The most essential element of the
definition is the use  of tensor products of  two $so(4)$ coset
representations. This is necessary for interpretation as generalized
spinor and for proper identification of $so(3,1)$ Lorentz algebra.

\vm

 To sum up,  all what is needed to obtain Kac Moody counterpart  of local
 $ so(3,1)$ allowing interpretation in terms of generalized spinors is the
replacement

\begin{eqnarray}
F^{1,k}_1&\rightarrow& iF^{1,k}_1
\end{eqnarray}

\noindent in the  tensor product of two $so(4)$ representations and
 in the definition of Super Virasoro algebra.  The task is to show that
super conformal invariance is not lost  or equivalently:  it is possible
to replace the generators $F^{1k}_1$ with $iF^{1k}_1$ in the
representation of Super Virasoro generators without spoiling the
algebra.   The definition of super generators $G^k$ allows this as the
following argument shows. \\ a) Isospin generators in  sector where
 $F^{am}$ act are of form $T^a \propto f_{abc}F^{bm} F^{cn}$.  This means
the following transformation rule for isospin generators

\begin{eqnarray}
(T^1,T^2,T^3)_1\rightarrow (T^1, iT^2,iT^3)_1
\end{eqnarray}

\noindent in replacement $F^1 \rightarrow iF^1$.  Same replacement
 must be done also in second $su(2)$ sector of $so(4)$ associated with the
sector $i=1$.\\ b)   Super generators involve terms  of general form $
f_{abc}F^{ak}F^{bl}F^{cm}$ from the $su(2)$ sector where $F^{am}$ act.
  This implies that $G^k$ transforms as $G^k \rightarrow iG^k$ in the
replacement
 $F^1 \rightarrow iF^1$.  If one modifies the definition of $G^k$ by
 $G^k\rightarrow iG^k$ Super Virasoro algebra suffers no change.

\vm

The states in Super Virasoro representation can  be labeled by discrete
eigenvalue of the axial Lorentz generator $\Sigma^{03}$ and the
representation contains infinite number of different eigenstates of this
generator.
 An interesting problem is the relationship of the coset representation to
the known discrete series unitary representations of Lorentz group,  which
are also infinite dimensional \cite{Gelfand}.

\subsubsection{Kac Moody counterparts of quarks leptons and gauge bosons }

A nontrivial problem is related to the construction of the exact
 Kac Moody counterparts of the imbedding space spinors and gauge bosons.
Consider first the construction in case of  imbedding space spinor.  \\ a)
The ground states for an appropriate tensor product of Kac Moody
representations associated with $M^4$ and $SO(4)$ plus $u(1)$  and
$su(3)$  degrees of freedom should behave as  imbedding space spinor and
different chiralities of this spinor correspond to quarks and leptons of
single generation.  \\ b) $M^4$ and $so(4)$ degrees can be treated in
identical manner using spin or isospin $1/2$  Ramond representations
having $ (c,h)=(0,0)$.  Single Ramond representation gives single spin or
isospin doublet. This means that one must take tensor product of two
Ramond representations in both $M^4$ and $CP_2$ degrees of freedom to
obtain the Kac Moody counter part of H-spinor having $4\times 4$
components. These representations correspond in natural manner to the two
possible representations of $su(2)_R\times su(2)_L$ obtained as tensor
products  $(k_R=0,\lambda=0)\times (2,R)$ and $(2,R)\times
(k_R=0,\lambda=0)$

\begin{eqnarray}
H-spinor &\leftrightarrow& D(M^4) \otimes D(CP_2)\nonumber\\
D(M^4)=D(CP_2)&=& ((k_R=0,\lambda=0)\otimes (2,R) )\times
 ((2,R)\otimes (k_R=0,\lambda=0)) \nonumber\\
\
\end{eqnarray}

\noindent The number of the vacuum states is correct and one can define
various chirality operators associated with $M^4$ and $CP_2$ degrees of
freedom and identify leptons and quarks as states of different
H-chirality. Super fields $\Gamma_i$ associated with the $so(4)$  and
$so(3,1)$ super Virasoro representations change leptonic to quark
chirality and vice versa. \\
 c) The treatment of $U(1)$ degrees of freedom is simple:  one just adds
 the appropriate  Ramond type $U(1)$ Kac Moody representation to lepton and
quark like representations as a tensor factor and takes into account that
 K\"ahler charges have their coorect values  for quarks, leptons and their
antiparticles.  The treatment of color degrees of freedom is similar: the
tensor factor associated with leptons and quarks is $N-S$ and Ramond
representation of color Super Virasoro respectively.

\vm

The treatment of bosonic sector proceeds in similar manner. \\
 Also in NS- sector one must use the tensor product of two N-S
 representations in both $M^4$ and  $CP_2$ sectors.

\begin{eqnarray}
Boson &\leftrightarrow& D(M^4) \otimes D(CP_2)\nonumber\\
D(M^4)=D(CP_2)&=& ((k_R=0,\lambda=0)\otimes (2,N-S)_i )\otimes
 ((2,N-S)_i\otimes (k_R=0,\lambda=0)) \nonumber\\
\
\end{eqnarray}

\noindent Dirac equation can be generalized so that it applies to both
Ramond and N-S case.

\subsection{ Super Virasoro in  $U(1)$ degrees of freedom}

  The treatment of the  $U(1)$ group associated with   K\"ahler charge
(essentially electroweak hyper charge)  is  completely analogous to
 the representation associated with $M^4$ degrees of freedom in string
models.   Nontrivial problem is
 associated with the nonvanishing vacuum expectation value

\begin{eqnarray}
h(U(1)) &= &\frac{p^2}{2}=\frac{Q_K^2}{2k(U1)}
\end{eqnarray}

\noindent  of $L_0$.  $p$ can be identified as  the 'anomalous' K\"ahler
 charge of the particle deriving from the additional $U(1)$ term appearing
in $CP_2$ spinor connection and making $CP_2$ spinor structure possible.
One has $Q_K=1/3$ for quarks and $Q_K=-1$ for leptons.

\vm

 The requirement that $su(2)\times u(1)$ acts as symmetry group means that
 one can mix the generators of $I_3$ and $Q_K$. This requires that the
values of  $u(1)$ central charge is
 and  vectorial central charge  are identical:

\begin{eqnarray}
2k(U(1))&=&m=k_R+2
\end{eqnarray}

\noindent  The dependence of $k(U(1))$ on $m$ introduces additional
conribution   into  the mass formulas. At the symmetry limit  one has
$h(U(1),L)=1/2$ for leptons.

\vm

For quarks one has at symmetry limit $h(U(1),q)= 1/18$.  Free  quarks have
necessarily mass of order Planck mass unless the vacuum weight $h_0$ for
quark representations   contains a compensating term.   One might think
that here lies an explanation for the confinement of fractional charges.
It however turns out that quarks must be
 massless: this of course is also the only physically acceptable
alternative in view of successes
 of parton model.

\vm

 In  string model $c_{tot}=0$ is important consistency condition and this
might be the case also now although it is difficult to see why:  Poincare
and color invariances are not lost and $so(4)\times u(1)$ is not an actual
symmetry.   In present case $c=0$ for  $M^4$ Super Virasoro,  for $m=2$
$so(4)$ Super Virasoro and for $su(3)$ Super Virasoro  (as will be found)
but   for $U(1)$ Super Virasoro $c=3/2$.  It seems however impossible to
apply coset construction  (by applying it to $so(4)\times u(1)$ and
$su(2)\times u(1)$)  in order to get vanishing value of $c(U(1))$ without
eliminating $U(1)$ contribution to
 Virasoro generators totally and at same time loosing nice explanation for
the experimental absence of fractionally charged states.

\subsection{Super conformal invariance and color symmetry}

The construction of  a unitary representation of Super Virasoro algebra
 having $(c,h)=(0,0)$ based on color super Kac Moody group proceeds along
the following lines.

\vm

a) Color is unbroken gauge symmetry:  this suggests that the
representation  in question corresponds to vanishing Kac Moody central
charge $k_B$. $k_B=0$
 implies  the vanishing of the conformal central charge $c$.  The vacuum
expectation value $h$ of $L_0$ should also vanish. $(c,h)=(0,0)$
representation of Virasoro has extension to a unitary Super Virasoro
representation so that there are good hopes in case of $su(3)$,  too.

\vm

b) Super Virasoro representations are of Ramond or Neveu-Schwartz type and
 this raises a problem.  Ordinary Ramond representation has spinorial
vacuum state and in case of color group the vacuum would correspond to a
spinor of 8-dimensional Euclidean space.  This is certainly an unphysical
property. The Kac Moody algebra and/or its super counterpart must  differ
from the  ordinary Kac Moody algebra in some crucial manner.

\vm

c)  Color Lie-algebra  is unique among other Lie-algebras in that  it
 can be extended to a super algebra.  The point is that for triplet
representation color generators allow also anticommutator structure

\begin{eqnarray}
Anti(T^a,T^b)\equiv T^aT^b+T^bT^a&=& k \delta (a,b) I +d_{abc} T^c
\end{eqnarray}

\noindent  For color triplet representation one has $k=1/3$ and the
 quantities $d_{abc}$ are completely symmetric 'structure constants'
appearing in the completely symmetric color singlet constructed from three
color  octets.  This  implies that color algebra can be extended to super
algebra
 $\{T ^a,F^a\}$:

\begin{eqnarray}
Comm(T^a,T^b) \equiv T^aT^b-T^bT^a&=& f_{abc}T^c\nonumber\\
Comm(T^a,F^b) &=& f_{abc}F^c\nonumber\\
Anti(F^a,F^b)\equiv F^aF^b+F^b F^a&=& k \delta (a,b) I +d_{abc} T^c
\end{eqnarray}

\noindent Super Jacobi identities

\begin{eqnarray}
Comm(Anti(F^a,F^b),T^c)- Comm(Comm( F^b,T^c), F^a) +
Anti(Comm(T^c,F^a),F^b)&=&0 \nonumber\\
\
\end{eqnarray}

\noindent boil down to  the identities

\begin{eqnarray}
d_{abd} f_{dce}-f_{bcd}d_{dae}+f_{cad}d_{dbe}&=&0
\end{eqnarray}

\noindent These identities follow from the condition

\begin{eqnarray}
Comm(T^d,d_{abc}T^aT^bT^c)&=&0
\end{eqnarray}

\noindent  The simplest representations of super color algebra are
representations for which the condition

\begin{eqnarray}
T^a &=&F^a
\end{eqnarray}

\noindent holds true.  The unique feature of these representations is the
 absence of additional degrees of freedom characteristic to the ordinary
super algebras. It is just this property, which suggests that the central
term of color  Super Kac Moody should be modified to include the $d_{abc}$
term.

 \vm

e) The next step is to construct Color Super Kac Moody algebra
 $\{T^a_n,F^a_n\}$.  The definition of this algebra is quite uniquely
fixed by dimensional requirements. The definining equations are the
following ones:

\begin{eqnarray}
Comm(T^a_n,T^b_n) &=& f_{abc} T^c_{m+n} +
k_B (n-m)\delta (a,b) \delta (m+n)I\nonumber\\
Comm(T^a_n,F^b_n) &=& f_{abc} F^c_{m+n} \nonumber\\
Anti(F^a_n,F^b_n) &=& (k\delta (a,b)+ d_{abc}T^c )\delta (m+n)\nonumber\\
\
\end{eqnarray}

\noindent  For fermionic generators  $n$ is either integer (Ramond
 representation) or half integer (N-S representation).  The condition

\begin{eqnarray}
F^a_0&=& T^a
\end{eqnarray}

\noindent implies that the physically interesting  Ramond representations
does not have the usual $d=8$ spinor degeneracy. The value of the fermionic
central charge is fixed by vacuum representation. For quark triplet one has
$k=1/3$.

\vm

f) Concerning the construction of Super Virasoro representation there are
two clues. The first  and probably wrong clue (as it turned out) is
 provided by the observation that  coset construction based on groups
$su(3)$, $ u(2)$  with central charge $k_B=1/2$ gives $c=0$ representation
of Virasoro
 algebra as is clear from the general formula for $c$ in coset construction
 \cite{Goddard,Bible} given by

\begin{eqnarray}
c&=& c(su(3))-c(su(2))-c(u(1))\nonumber\\
c(g)&=& \frac{2k dimg}{2k+ c_g}\nonumber\\
c_g&=& \frac{f^{abc}f_{abc}}{dimg}
\end{eqnarray}

\noindent  One has $c_g=3,2$ and $0$ for $g=su(3), su(2)$ and $u(1)$
respectively so that c indeed vanishes for $k= 1/2$.  The representation in
question has $h=0$ as one finds easily.   I have not however been able to
find the extension of this representation to Super Virasoro
representation.  The nonvanishing value of Kac Moody central charge $k$ as
well as symmetry breaking to $u(2)$  (Virasoro commutes with $u(2)$ Kac
Moody) suggests that this representation is not physically interesting:
color group  should act as unbroken  local gauge group.

\vm

 g) The second clue comes from the observation that  ordinary color Super
Kac
 Moody allows purely 'fermionic' representation \cite{Goddard,Bible}
 with bosonic generators defined in terms of the fermionic generators as

\begin{eqnarray}
T^a_n&=& K :f_{abc}\sum_{m\neq 0,m\neq n} F^b_{n-m}F^c_m:
\end{eqnarray}
\noindent
Note the exclusion of zero modes in present case.   The value of the
constant $K$ is $K=1/2$ for the ordinary Kac Moody  algebra.  The value of
the bosonic central charge is $k_B= c_g/2=3/2\neq 1/2$  so that this
representation cannot be used in the previous coset construction although
the specific form of the generators suggests the
   possibility  to define super Virasoro generators $G$ as linear
combinations of generators of form $G_g\propto \sum_{a \in g}T_aF^a$, $
g=su(3),su(2),u(1)$.

\vm

A  suitable modification of this representation might  make sense also in
the
 present case since the general structure of the commutation relations is
correct. \\ i) The value of constant $K$ can be fixed by studying the
contribution of $F^a_0 = T^a $ to  the c-number part of $T^a_0$ and by
requiring that this reduces to the ordinary Lie-algebra generator  $T^a$:
this  condition reads as $T^a =K f^{abc} T_bT_c$ and gives

\begin{eqnarray}
K&=& \frac{2}{3}
\end{eqnarray}

\noindent ii) The commutator $Comm(T^a_n,T^b_m)$ involves besides the
ordinary
 term given by $\frac{16}{9} f_{abc}T^c_{n+m}$  terms  proportional to
$d_{abc}$ and involving $T^c$ besides the  two fermionic generators.
$T_c$ can  be eliminated using the commutation relations of the  bosonic
generators.  Due to the   identity

\begin{eqnarray}
d_{abc}(f_{ACc}f_{bBd}f_{dDa}
+f_{ADc}f_{bCd}f_{dBa})
&=& \frac{5}{24} f_{ABe}f_{eCD}
\end{eqnarray}

\noindent relating the structure constants $f$ and $d$ this contribution
is just the needed  compensating term $-\frac{7}{9} f_{abc}T^c$.
 The identity follows by considering the commutator $Comm(T^a,T^b)$
 for super color algebra in cm degrees of freedom,  using the
representation $T^a= \frac{2}{3} f_{abc}F^bF^c$ and expressing the
commutator in terms of anticommutors  only. \\ iii)   For the ordinary
representation  the value of the Kac Moody central charge would be equal
to $k_B=3/2$ but in present case also the $d_{abc}$ terms give
contribution to the central charge term and for $k=1/3$ representation the
terms cancel each other! This can be understood by noticing that central
charge term comes from two anticommutators of  the 'fermionic' generators
in the commutator.  For the ordinary representation the term is
proportional to $ C_{ad}= f^{abc}f_{dbc}=\frac{3}{2} \delta (a,d)$. In
present case one has more general formula

\begin{eqnarray}
C_{ad}&=& f^{abc}D_{be}D_{cf}f^{def }\nonumber\\
D_{ab}&=& k \delta {ab} +d_{abc} T^c_0
\end{eqnarray}

\noindent The  terms  proportional to $T^c$  must vanish  by the $su(3)$
 invariance of the central term.  The terms quadratic in $d_{abc}$ give a
contribution, which must be proportional to unit matrix and a
straightforward calculation for $a=d=Y$ shows that the two contributions
cancel each other exactly.  The vanishing of $k_B$ implies that if Super
Virasoro for the representation in question exists it  has  $c=0$.

\vm

h)  The simplest guess is that Virasoro generators $L_n$  are given
 by the standard Sugawara construction

\begin{eqnarray}
L_n&=& \frac{1}{\beta}: \sum_mT^a_{n-m} T^a_m:\nonumber\\
\beta&=& 2k_B+ c_g= 3 \ for \  g= su(3)
\end{eqnarray}

\noindent Since $k_B=0$ the normal ordering is not in fact necessary.  The
 general form of the super generator $G_n$  should be $T_aF_a$:  this is
achieved takin $G_n$ to be of the following general form

\begin{eqnarray}
G_n&=& g f_{abc}\sum_{k,lm\neq 0} F^a_kF^b_l F^c_m \delta (n-k-l-m)
\end{eqnarray}

\noindent Again the normal ordering is unnecessary. The terms proportional
 to $d_{abc}$ in the anticommutator

\begin{eqnarray}
L_{n+m}= \frac{1}{2}Anti(G_n,G_m)
\end{eqnarray}

\noindent  vanish by Super Jacobi identities and one gets the standard
Sugawara form  by choosing the constant $g$ to have the value given by the
condition

\begin{eqnarray}
9g^2&=& \frac{1}{\beta} K^2= \frac{4}{27}
\end{eqnarray}

\noindent giving $g= \frac{2}{9\sqrt{3}}$. The commutators $Comm(L_n,L_m
)$  are certainly of correct form  and  since  they are expressible
using the  commutators between Virasoro generators and super generators,
it is very plausible that also the commutors $Comm(L_n,G_m)$ have  correct
form.

\vm

To sum up, the unique property of $su(3)$ Lie-algebra allowing
anticommutator
 structure allows to construct Super Virasoro algebra associated with color
group with physically acceptable properties in Ramond sector. This is an
additional item in the long list of very special properties of $CP_2$
geometry making the choice $H=M^4_+ \times CP_2$ as imbedding space
unique.

\subsection{How to achieve Super Virasoro invariance in leptonic case?}

In leptonic case generalized spinor is tensor product of Ramond type
 representations with N-S type representation in color degrees of freedom.
This leads to problems with Super Virasoro invariance. One should somehow
be able to transform the half odd integer Super Virasoro generators
associated  with $su(3)$ degrees of freedom to generators labeled by
integer.  It can be assumed that the super Virasoro generators $G^n,
n=0,1,2,..$ associated with $so..\times u(1)$ degrees of freedom commute
with the generators $G^{k}(su(3)),k= \pm 1/2,...$ associated with color
degrees of freedom.  It can be assumed that the super Virasoro generators
$G^n, n=0,1,2,..$ associated with $so..\times u(1)$ degrees of freedom
commute with the generators $G^{k}(su(3)),k= \pm 1/2,...$ associated with
color degrees of freedom.

\vm

 A possible resolution of the problem is based on the following
observations: \\ a)  There must be operator
 $B^{1/2}$ of conformal weight $\Delta=1/2$  acting on $u(1)$ degrees of
freedom since K\"ahler charge contributes to the vacuum weight  by $\Delta
h=1/2$. Assume therefore  the existence of operators $B^{1/2}$ and
$B^{-1/2}$, which are hermitian conjugates and  anticommute to unit matrix
whereas their commutator vanishes:

\begin{eqnarray}
Anti(B^{1/2},B^{-1/2})&=&1\nonumber\\
Comm(B^{1/2},B^{-1/2})&=&0
\end{eqnarray}

\noindent  The conditions imply $B^{1/2}B^{-1/2}=1/2$. Any diagonal
matrix,
 whose diagonal elements are of form $ exp(i\phi)/\sqrt{2}$ gives a
realization of these relations so that there are good hopes of finding
solution to the conditions.  These operators anticommute with Ramond type
super generators and commute with all color Super Virasoro generators.  \\
b) The commutation relations between Virasoro generators and Super
generators do not produce troubles if one multiplies color Super
generators $G^{n+1/2}$ , $n>0$ with $B^{-1/2}$  and $G^{n-1/2}$, $n<0$
with $B^{1/2}$ to obtain color contribution to new Ramond type generator
$\hat{G}^{n}$,$n\neq 0$.  For $n=0$ special care is needed since both
$B^{-1/2}G^{1/2}$ and $G^{1/2}G^{-1/2}$ contribute to $\hat{G}^0$.  A
linear combination of these terms is obviously needed so that one has:

\begin{eqnarray}
\hat{G}^n&=& G^n(so..\times u(1)+ B^{1/2}G^{n-1/2}(su(3), \ n>0 \nonumber\\
\hat{G}^n&=& G^n(so..\times u(1)+ B^{-1/2}G^{n+1/2}(su(3), \ n<0 \nonumber\\
\hat{G}^0&=& G^0(so..\times u(1))+
k(B^{-1/2}G^{1/2}(su(3))+B^{1/2}G^{-1/2}(su(3))) \nonumber\\
\
\end{eqnarray}

\noindent where $k$ must be chosen so that $G^0G^0=L^0$ condition holds
true.   \\ c) The only problems are related to the central term in the
anticommutator

\begin{eqnarray}
Anti(\hat{G}^m,\hat{G}^n)&=& 2L^{m+n}+ \frac{c}{3}(n^2-\frac{1}{4})\delta
(m,n)
\end{eqnarray}

\noindent since color contribution to the anticommutator is of form
 $(n+1/2)^2-1/4$ rather than $n^2-1/4$.  No troubles are  however
encountered if color algebra has vanishing central charge as it indeed
does!

\section{Identification of elementary particles as representations of
 p-adic Super Virasoro}

The  concept of generalized H-spinor implies unique identification of
 elementary particles as ground states of Super Virasoro representations.

\vm

\noindent a) Massless  elementary    particles   are identified as
generalized
 H-spinors having $(c,h)=(0,0)$.   Only $(J=1/2,I=1/2)$ fermions are
obtained and sfermions typical for
 super symmetric theories  are absent.   Bosons have $(J=1/0,I=1/0)$ and
again exotics are absent.  Absent is also Higgs doublet. \\ b) All
elementary particles   correspond to Super Virasoro representation of
$U(1)$ group with  $(c,h)= (3/2, Q_K^2/2)$.   For fermions the
reprensentation is of Ramond type and for bosons of type N-S.\\ c)  With
respect to color group leptons and bosons are N-S representations and
quarks correspond  to Ramond representations.

\vm

 Physical states are either  annihilated by Super Virasoro generators
 $L_n,n\neq 0$ and $G_n,n\neq 0,1/2$ or these operators create zero norm
states.  Same applies to the generators $T_n$ of the vectorial $su(2)$
subgroups  associated with $so(3,1)$ and $so(4)$: otherwise one would get
enormous mass degeneracy since $SO(4)$ Super Virasoro commutes with
vectorial $su(2)$ Kac Moody generators. The assumption  is in accordance
with conserved vector current hypothesis.   Color Kac Moody acts as an
ordinary local gauge group  and one must require that same conditions
apply to the generators of color Kac Moody.   A representation of local
Lorentz group can be constructed in terms of 'fermionic' generators of
$M^4$ Virasoro.  Since conformal anomaly vanishes in $M^4$ sector this Kac
Moody algebra has vanishing central charge and one can say that both local
$so(4) $   local $SO(3,1)$ acts as a local
 gauge group and  physical states satisfy appropriate conditions.
Symmetry breaking implies
 that these groups reduces to local $su(2)_V$.

\vm

In TGD quarks and leptons correspond to spinors of opposite H-chirality
defined as product of $M^4$ and $CP_2$ chiralities and  $B $ and $L$ are
exactly conserved quantum numbers.  One can associate definite
 baryon and lepton numbers  both to bosons and fermions  also in present
approach \cite{TGD}.   Super Virasoro generators with half odd integer
conformal weight act as operators transforming quarks into leptons and
vice versa.    Leptoquarks, that is bosons with lepton and baryon number,
if they exist,   turn out to  have
 Planck mass.   Exact conservation of $B$ and $L$ implies that only
$\Delta \in Z$ excitations are  allowed and thermal mixing of different
G-parities  is not possible.

\vm

Thermal mixing  between different G-parities, if possible,  would imply
breaking of $H$-chirality   analogous to the ordinary breaking of
Minkowski space chirality associated with
 massivation and provides an obvious explanation  for  matter antimatter
asymmetry:  symmetry
 breaking effects would be  certainly small, of order $O(p^2)$ at the
level of matrix elements:  naive  order of magnitude estimate for lifetime
of proton as $M_p/p^2$ for $p= M_{107}$ gives lifetime of order $10^{48}$
years.  In TGD there is  however a mechanism for generation
 of matter antimatter asymmetry based on  different rates for topological
evaporation of fermions and antifermions in  (necessary) presence of
vacuum K\"ahler electric fields.

\vm

A more stringent requirement is suggested by string models:  allow only
excitations with  G-parity equal to $+1$ or $-1$.   This condition
eliminates leptoquarks from spectrum and the excitations of leptons to
quarks with wrong color triality $t=0$ instead of $t=1$.
 In TGD  the absence of actual tachyons in TGD   forces  G-parity rule.

\vm

The main differences between p-adic TGD and more standard theories are
 following: \\ a)  Higgs doublet is absent in TGD.\\ b) Leptoquarks are
absent.\\ c) Superpartners of ordinary elementary particles are absent. \\
d) Graviton is not obtained in the simplest scenario as  a massless
elementary
 particle.  States with quantum numbers of graviton are obtained as tensor
products of basic representations but these  correspond to ordinary hadrons
rather than actual graviton.   The result is in accordance with the basic
assumption of p-adic approach:  p-adic super conformal invariance makes
sense only in the approximation that space time is flat \cite{padTGD}.
There is analogy with string model:  in string model graviton corresponds
to closed string, which corresponds to closed 3-surface without boundary.
In p-adic approach \cite{padTGD} one however considers  only  surfaces,
which are essentially pieces of Minkowski space and have   outer boundary.
Perhaps the absorption  of graviton corresponds to  topological sum of
absorbing particle and closed 3-surface describing graviton.  One can
consider   the possibility that gravitons correspond to two-sheeted
surfaces obtained by gluing to essentially flat pieces of $M^4$ together
along their boundaries:  this would imply a formal generalization of
string model and graviton would be obtained.   This amounts to forming  a
tensor product of two Kac Moody spinors but it is
 not evident whether thermal massivation allows any spin 2 massless
states. \\
 c) Color excitations of  leptons obtained using the generators of N-S
type  color algebra  with triality $t=1$ are possible. It turns out that
massless excitations exist  for leptons as well as quarks.   Color bound
states of colored excitations are
 therefore possible and  a conformation for the prediction  of  new branch
of physics is obtained  \cite{Lepto,Heavy}.  This prediction is in
accordance with the TGD inspired explanation
 \cite{Lepto,Heavy} of anomalous $e^+e^-$ events observed in heavy ion
collisions in terms of leptopions, bound  states of color excited leptons.
In the fourth  paper of the  series the rather deep consequences
 of colored  excited leptons and quarks are discussed.

\subsection{Definition of fermionic Dirac operator  and the problem of
tachyonic
 ground states}

One should generalize Dirac equation in such a manner that mass shell
condition
 results as a consequence of this equation.  The generalization should
apply in both fermionic and bosonic case and in bosonic case it should
give the usual orthogonality condition for polarization vector and four
momentum vector.   Both bosons and fermions correspond to Kac Moody
counterparts of H-spinors and the  only difference between bosons and
fermions is that they correspond to different representations of Super
Virasoro.

\vm

In order to construct Dirac operator one needs gamma matrices.
 For fermions one can define Minkowski space gamma matrices in terms of
super generators  $F^{a,K}_i$, $i=1,2$. One has  $K=0$ in fermionic case
and
 $K=1/2$ in bosonic case.

\begin{eqnarray}
\gamma^0&=&iF^{1,K}_1 \  \ \gamma^1= F^{2,K}_1\nonumber\\
\gamma^2&=& F^{1,K}_2\  \ \gamma^3=F^{2,K}_2
\end{eqnarray}

  \noindent  Here the representation associated with $so(3,1)$ is regarded
 as $so(4)/su(2)$  coset representation and this explains the imaginary
unit.  For bosonic gamma matrices one can define conjugate matrices  in
terms of $F^{i,-K}_k$. It is also possible to define $CP_2$ type gamma
matrices applying previous formulas in $so(4)$ degrees of freedom and
dropping the imaginary unit.  Generalized Sigma matrices can be
constructed in standard  manner as commutators of gamma matrices.

\vm

There are infinitely many gauge equivalent ways to define gamma matrices by
performing $su(2)$ rotations in both $so(4)$ factors of the tensor product
defining the representation. These rotations are pure gauge rotations and
cannot affect physics. In particular, the additional degeneracy of states
resulting from the gauge invariance has no physical consequences and
introduces just an  integer factor in partition function.

 \subsection{Fermionic Dirac equation}

Ground state should have such mass that for both  for neutrinos, D type quarks
and bosons  $L^0=5/2$  states  are massless whereas for charged leptons and U
 type quarks $L_0=3/2$ states are massless.  This requirement can be
contrasted with the anticommutation relations for  Ramond type Super
Virasoro generators.   To begin with,   the representation of $G^0(tot)$
given by

\begin{eqnarray}
G^0(tot)&= &\sqrt{\frac{2}{k(F)}}p^k \gamma_k +G^0
\end{eqnarray}

\noindent where $k(F)$  scale factor, whose value turns out to be
 $k(F)=3/2$ for purely physical reasons. Anticommutations for super
generators  read as

\begin{eqnarray}
Anti(G^m,G^{-n})&=&  -\frac{1}{k(F)}p^2 \delta (m,n) +L^{m-n}
- \frac{c}{3}(m^2-\frac{1}{4})
\delta (m,n)\nonumber\\
\
\end{eqnarray}

\noindent $G^0G^0=0$, which can be regarded as a  generalization for the
 square of
 massless Dirac equation   gives mass formula

\begin{eqnarray}
p^2&=& k(F)(L^0- \frac{c}{24})
\end{eqnarray}

\noindent  The should be  ground state with $L^0 = 5/2$  is  massive and
possesses Planck mass so that something goes wrong with the generalization
of the massless Dirac equation.  For ground state ($L^0=0$ ) one has $m^2=
-3/32$ for $c=3/2$.  One  might regard this state as tachyon but in p-adic
case the square root of $-3/32$ exists as real p-adic number under rather
general conditions:  for instance for  Mersenne primes different from
$M_2$.

\vm

One can modify the generalized massless Dirac equation for
 fermions by introducing fermion mass so that correct massless states are
obtained. Symmetry breaking actually requires that $M^4$ chiralities get
mixed so that massless Dirac equation is not realistic and mass term is
needed. The problem is that the addition of scalar mass term  breaks
chirality invariance by
mixing different H-chiralities.  The properties of induced spinors suggest a
resolution
of the problem. The $CP_2$ part for the  trace of second fundamental for for a
submanifold appears
as mass term in Dirac equation for induced spinors. Originally the trace of
 of the second fundamental form was identified as  Higgs field.   This
term mixes $M^4$  chiralities but respects H-chirality conservation.  In
 present
case the mass operator must be linear combination of gamma matrices in $so(4)$
degrees of freedom and must anticommute with the operator $G_0$. This is
achieved if  Higgs term is linear combination of gamma matrices belowing to
$su(2)_V$ since  fermionic Super Virasoro generators in coset representation
indeed anticommute with corresponding operators for factor group.
Only $\gamma$
matrices $\gamma_0$  and $\gamma_3$ commuting with the sigma-matrix
$\Sigma_{12}$ appearing in electromagnetic charge matrix are allowed.   Since
 the
values of $M^2$ are different for different charge states of fermions the
mass operator must be decomposed into a sum of parts operating in different
charged states only.

\begin{eqnarray}
M_{op}&=& (a\gamma_0+ b\gamma_3) P_+ +
(c\gamma_0+d\gamma_3)P_-\nonumber\\
P_{\pm}&=& \frac{1\pm 2I_3}{2}
\end{eqnarray}

\noindent The gamma matrices correspond to $CP_2$ tangent space gamma
 matrices
identified previously as $so(4)$ super generators.

\vm

  The generalization of Dirac
equation reads as

\begin{eqnarray}
(\frac{\sqrt{2}}{k(F)}(p^k\gamma_k+M_{op}) + G^0 )\vert PHYS \rangle &=&0
\end{eqnarray}

\noindent The solution to this equation is obtained in standard manner

\begin{eqnarray}
\vert PHYS\rangle &=& (\frac{\sqrt{2}}{k(F)}( p^k\gamma_k -M_{op})+ G^0 )\vert
phys \rangle \end{eqnarray}

\noindent  where $\vert phys\rangle $ satisfies Super Virasoro gauge
conditions and leads to the mass shell condition

\begin{eqnarray}
p^2&=&k(F)(L_0-\frac{1}{16})+M^2_{op}\nonumber\\
M^2_{op}&=& (a^2+ b^2)P_+ +(c^2+d^2)P_-
\end{eqnarray}

\noindent   The requirement that $L^0=2$ ($L_0=1$)  state is massless for
neutrinos and D type quarks (charged leptons and $U$ type quarks) gives
constraints on the  eigenvalues of the mass squared operator

\begin{eqnarray}
a^2+ b^2  &=& -k(F)(   2- \frac{1}{16})\equiv A_+= -\frac{3}{32}\cdot 31
\nonumber\\
c^2+d^2  &=& -k(F)(   1-
\frac{1}{16}) \equiv A_-=-\frac{3}{32}\cdot 15 \nonumber\\
k(F)&=& \frac{3}{2}
\end{eqnarray}

\noindent   From these equations one can solve $a,b,c$ and $d$.

\vm

The  solution ansatz relies on the existence/nonexistence of
square roots for $A_+,A_- $.  $\sqrt{A_-}$ does not exist
as p-adically real number ($\sqrt{p}$ is regarded real in this context) for
Mersenne primes except $M_{31}$. For $M_n \neq M_2= 3$  the existence of
 square
root   reduces to the existence of $\sqrt{31}$ in case of $A_+$.
The existence or nonexistence of square root is verified  easily by applying
repeatedly  resiprocity lemma stating that  for two primes $ p \  mod \  4=3$
 and
$q \ mod \ 4=3$  one has $p= x^2 \ mod \  q$
iff $q\neq x^2 \ mod \ p$.  Otherwise one has
 $p= x^2 \ mod \  q$ iff $q= x^2 \ mod \ p$.  A useful auxiliary result is the
existence of  $\sqrt{-3}$ as p-adically real number for all Mersenne primes
except $M_2=3$ since neither $-1$ nor $3$ allows p-adically real square root
for
Mersenne primes. This means that although the arguments of square roots are
negative as rational numbers their square roots can exist p-adically and
negative mass squared ground states are not actually tachyons.  $\sqrt{5}$
and
$\sqrt{2}$ exist for all Mersenne primes  except
 $M_2=3$ ($2=x^2 \ mod \ p$ for all $p \ mod \ 8=\pm 1$).
 The following table summarizes the situation
for different physically interesting  Mersenne primes.

\vl

\begin{tabular}{||c|c|c|c|c|c|c|c|c|c|c|c||}
 \hline\hline
$n$  &127 & 107&89&61&31&19&17&13&7&3&2 \\ \cline{1-12}\hline
$A_+$&  Y &Y   &Y &N &N &Y &Y &N &Y&N&N     \\ \hline
$A_-$&  N &N   &N &N &N &N &N &N &N&Y&N      \\ \hline \hline
\end{tabular}

\vl

Table. The existence of square root of $A_+$ and  of $A_-$ for different
Mersenne primes $M_n$ (Y=yes,N=no).

\vm

For $I_3=-1/2$ states the vector $(c,d)$ cannot
allow real unit vector except for $M_2=3$ . Perhaps the simplest solution to
the
condition $c^2+d^2=A_-$ is

\begin{eqnarray}
c&=& \sqrt{3A_-}\nonumber\\
d&=& \sqrt{-2A_-}
\end{eqnarray}

\noindent and makes sense for $M_n$, $n\neq 2$ ($M_2=3$).    Both square
roots exists since $3$ and $-1$ and $A_-$ do not allow real square root
whereas $2$ allows square root for $M_n,n>2$.  For $n=2$ the ansatz

\begin{eqnarray}
(c,d)&=& \sqrt{A_-}(n_1,n_2)\nonumber\\
n_kn^k&=&1
\end{eqnarray}

\noindent works.  The simplest rational solution is $(n_1,n_2)=(1,0)$ or
$(0,1)$.

\vm

For $I_3=1/2$ states similar ansatz works for $(a,b)$ in $(N,N)$ case. For
$(Y,N)$ case the vector $(a,b)$ allows unit vector and is of the form

\begin{eqnarray}
(a, b) &=& \sqrt{A_+}(n_1,n_2)\nonumber\\
n_kn^k&=&1\nonumber
\end{eqnarray}

\noindent Any Pythagorean triangle with sides smaller than $p$ defines
rational unit vector. For $n=2,3$ the only rational solution is $n_1=1$ or
$n_2=1$.

\subsection{Bosonic Dirac equation}

Also for bosons vacuum weight is $h=-5/2$ and massless states correspond
 to $L^0=5/2$. This necessitates the introduction of mass parameter to
bosonic Dirac equation.
 In bosonic case  the candidate for massless Dirac equation is

\begin{eqnarray}
G^{1/2}(tot)\vert PHYS\rangle &=&0\nonumber\\
G^{1/2}(tot)&=&\frac{ \sqrt{2}}{k(B)}p_k\gamma^{k}_{1/2}+G^{1/2}
\end{eqnarray}

\noindent where $k(B)$ is some parameter measuring the ratio of fermionic
and bosonic string tensions.   The equation leads to mass shell condition
mass shell condition  $p^2=k(B)L^0$, which  fails to give massless states
for $L^0= 5/2$.  In analogy with fermionic case 'Higgs' term to define mass
operator

\begin{eqnarray}
G^{1/2}(tot)&=&\frac{
\sqrt{2}}{k(B)}p_k\gamma^{k}_{1/2}+G^{1/2}+M_{op}\nonumber\\
M_{op}&=&
a\gamma_0^{1/2}+b\gamma_3{1/2}  \end{eqnarray}

\noindent The mass operator $M_{op}$ must be chosen so that $L^0=5/2$ states
are  massless.  The square of Dirac equation corresponds to $G^{-1/2}
G^{1/2}$ and gives mass shell condition

\begin{eqnarray}
p^2&=& k(B)L^0+M^2_{op}
\end{eqnarray}

\noindent    The construction also leads to a unique value of
$k(B)=3/2$.   In this case $M_{op}$ is determined from the condition

\begin{eqnarray}
M^2_{op}= a^2+b^2&=& -\frac{3}{4}\cdot 5\equiv A
\end{eqnarray}

\noindent  $\sqrt{5}$ and $\sqrt{-3}$ exist for all $M_n$ except
$M_2=3$. Thus the square root of the left hand side exists for all $n$ and
$(a,b)$  possessing unit vector defines acceptable  mass operator.

\vm

 The study of the bosonic sector shows that  the standard solution
ansatz for Dirac equation is not useful in bosonic case but that Dirac
equatiom  must be regarded as one  gauge condition allowing many other
solutions besides Dirac ansatz.   Bosonic Dirac equation  gives rise to
mass shell condition plus conditions
 stating that bosonic polarization and momentum vectors are orthogonal.
This follows from the fact that spin 1 bosonic state is proportional to
the operator

\begin{eqnarray}
\epsilon_k \gamma^k_{1/2}
\end{eqnarray}

\noindent and multiplication with $p_k\gamma^k_{-1/2}$ gives $
p^k\epsilon_k=0$.  Highest possible bosonic spin and isospin is $J=1,I=1$
as is clear from the fact that the operators $F^{1/2a}_i$, $i=1,2$ or
$i=3,4$ anticommute so that only $J=0,1$ and $I=0,1$ states result from
the tensor product.  The fact that graviton is absent is in accordance
with the assumption  about the approximate flatness of p-adic spacetime
surface, which is basic assumption  of p-adic conformal field theory limit.

\vm

To sum up,  the construction of fermionic and bosonic Dirac equation has
 provided two rather miraculous results:  purely p-adic mechanism for the
transformation  of ground state  tachyons  to Planck mass particles and
a  mild suggestion  that $M_{n}$ with $n=89,61,17$  are the only only
possible condensation level for gauge bosons among Mersenne primes.

\section{Description of Higgs mechanism as a breaking of super conformal
 symmetry}

In the following the phenomelogical description of Higgs mechanism  as
breaking of Super conformal symmetry is considered quantitatively.
General mass formulas are derived and the question what happens for mass in
condensation is considered. The considerations are somewhat out of date in
the sense that it is assumed that second order contribution to mass square
correspond to small quantum number limit in the sense that second order
term is of form $kp^2/64$ (Ramond) or $kp^2/16$ (N-S), where $k$ is small
integer. The  thermodynamical calculation of masses has shown that this
is  however not the case unless one performs suitable approximation or
unless the secondary condensation changes the situation.

\subsection{General mass formulas}

   Mass squared  for the  discrete series Super Virasoro representations
is  sum of $u(1)$ and $so(4)$ contributions:

 \begin{eqnarray}
M^2&=& M^2(U(1))+M^2(so(4))\nonumber\\
\end{eqnarray}

\noindent $so(4)$ contribution is given by the following general
expression  as vacuum weight $h$ for general discrete series Super
Virasoro representation:

\begin{eqnarray}
M^2(so(4))&=&h= \frac{( (m+2)P-mQ)^2 -4}{8m(m+2)} +
\frac{\epsilon}{16}\nonumber\\
m&=& k_R+2\nonumber\\
\epsilon&=&1(0) \ for \ Ramond (N-S)\nonumber\\
\
\end{eqnarray}

\noindent  Here one has $p=1,..,m-1$ and $q=1,..,m+1$. $p-q$ is odd for
Ramond representation and even for N-S representation. There are three
representations with vanishing mass squared $M^2=h=0$.  They have  $m=2$
and  therefore vanishing central charge  $k_R=0$ for the second $su(2)$
factor of $SO(4)$.   In Ramond sector one  $I=1/2 $  representation with
$(p,q)=(1,2)$ (leptons and quarks). In Neveu-Schwartz sector there are
$I=0$ representation with  $(p,q)=(1,1)$ (photon and gluons)  and $I=1$
representation with $(p,q)= (1,3)$ (intermediate gauge bosons $W$ and
$Z$). Higgs mechanism corresponds to breaking of super conformal symmetry,
when  Kac Moody algebra $su(2)_R$ and as a consequence  also super  super
conformal algebra develops central charge ($k_R \neq 0$).
 $I=1/2$ and $I=1$ states (quarks,  leptons, intermediate gauge bosons)
become
 massive in the breaking of conformal symmetry  $I=0$ states  (photon,
gluons) remain massless in the breaking of conformal symmetry.

\vm

$u(1)$ contribution is present for fermions only and for $Q_K=1$ states
 (leptons) it is given by the following  expression

\begin{eqnarray}
\frac{M^2(U(1))}{m_0^2}&=&  \frac{Q_K^2}{m}-\frac{1}{2}
\end{eqnarray}

 \noindent Exact conservation of electromagnetic charge
$Q_{em}= Q_K/2+I_3$ correlates $Q_K$ with $Q$.

\vm

$P,Q,$ and $ m$ deviate in order $O(p)$ from their values for massless
 representations.  One can expand the expressions for $P,Q$ and $m$ and
$h$ in power series
 with respect to $p$  and $O(p^3)=0$ approximation is extremely accurate:

\begin{eqnarray}
m&=& 2+m_1p +m_2p^2+..\nonumber\\
P&=&p_0+p_1 p+p_2p^2+...\nonumber\\
Q&=& q_0+q_1p+q_2p^2 +...\nonumber\\
Q_K&=&q_0^K+q^K_1p+q^K_2p^2+... \nonumber\\
\
\end{eqnarray}

\noindent Expanding $M^2(so(4))$  and $M^2(u(1))$ in
 power series one obtains the general mass formula

\begin{eqnarray}
M^2&\simeq& M_1^2+ M_2^2 \nonumber\\
\end{eqnarray}

\noindent For Ramond representation the  formula gives

\begin{eqnarray}
M_1^2& =& (3t+ q_K^1)p \nonumber\\
M_2^2&=& \frac{X}{64}p^2\nonumber\\
X&=& (8(2p_1^2+q_1^2-2p_1q_1)+ 32(q_1^K)^2-13m_2)\ mod \ 64\nonumber\\
m_1&=& 64 t
\end{eqnarray}

\noindent The requirement that lowest order contribution is not of  order
Planck mass gives the condition $m_1= 64 t$.   It should be noticed that
second order contribution to mass  depends only weakly on the values of the
integers appearing in it.

\vm

N-S representations can be assumed to have vanishing K\"ahler charge at
 elementary particle level.
 For N-S representations with $I_3=0$ one obtains

\begin{eqnarray}
M_2^2&=& \frac{X}{16}p^2\nonumber\\
X&=& (q_1^2-12km_1 +2(2p_2-q_2))\ mod \ 16\nonumber\\
2p_1-q_1&=& 8k
\end{eqnarray}

\noindent The condition $2p_1-q_1=8k$ follows from the requirement that
 lowest order contribution is small in Planck mass scale.

\vm

 For $I_3=1$ N-S representation the mass formula gives

\begin{eqnarray}
M_1^2&=& kp \nonumber\\
M_2^2&=& \frac{X}{16}p^2\nonumber\\
X&=& (q_1^2-12km_1 +2(m_2-2p_2+q_2))\ mod \ 16\nonumber\\
2p_1-q_1-m_1&=& 8k
\end{eqnarray}

\noindent Again the smallness of lowest order contribution gives
constraint on symmetry breaking parameters.

\vm

 Since K\"ahler charge is $Q_K^0=1/3$ for quarks the  $U(1)$ contribution
to mass squared is of order Planck mass

\begin{eqnarray}
M^2(u(1))&=& \frac{Q_K^2}{m}= M^2_0+M^2_1+M^2_2\nonumber\\
M_0^2&=& \frac{1}{18} \nonumber\\
M_1^2&=&  \frac{q_1^K}{3}p \nonumber\\
M_2^2&=& ( \frac{q_1^2}{6}-
\frac{m_2}{36})p^2\nonumber\\
\end{eqnarray}

\noindent unless vacuum weight for quark representations contains a
compensating contribution making massless states possible: this turns out
to be the case.  This is  physically as the only possibility since partons
are known to be essentially massless.  In thermal
 approach massless quarks  are predicted to spend fraction $\frac{1}{p}
\simeq 2^{-107}$ of time in Planck mass
 state.

\vm

 The relationship between $\Delta Q_K$  and $\Delta Q$  is fixed from the
 renormalization invariance of $Q_{em}$:

\begin{eqnarray}
q_1&=& \epsilon (L) q_1^K\nonumber\\
\epsilon (L)&=& 2I_3= \pm 1
\end{eqnarray}

\noindent  where the sign factor depends on the isospin of the lepton.
 Electroweak symmetry breaking results.  The actual breaking involves
however also different critical condensation level for lepton and its
neutrino so that one cannot deduce the ratio of electron and neutrino
masses from the formula.  It will be found that the sign of $q_1^K$ is
same for, say, proton and neutron.

\vm

For certain primes the real counter part of the $so(4)$ contribution  (and
also of $u(1)$ contribution given by the  canonical correspondence between
 p-adic and  real numbers is minimized.
 If $p$ satisfies the condition $p \ mod \ 64 =63$ so that $p$ can be
written
 in the form

\begin{eqnarray}
p &=& 2^n- 64k-1
\end{eqnarray}

\noindent the inverse of $64$ in the finite field $G(p)$ formed by p-adic
numbers mod  p is given by

\begin{eqnarray}
\frac{1}{64} \ mod \ p= 2^{n-6} + small  \ terms
\end{eqnarray}
\noindent Otherwise the inverse is

\begin{eqnarray}
\frac{1}{64}&=& 2^{n-1} + small \ terms
\end{eqnarray}

\noindent  Mersenne primes provide an example of the primes satisfying the
 condition. In \cite{padTGD} it was erroneously argued that $p \ mod  \ 4
=3$ guarantees that inverse satisfies the equation above.

\vm

 With above constraint on the allowed values of $p$  the contributions to
 mass are

 \begin{eqnarray}
M_1^2&=& \frac{(3k+ q_K^1)}{p} \nonumber\\
M_2^2&=& \frac{X}{64p}\nonumber\\
X&=& (8(2p_1^2+q_1^2-2p_1q_1)+ 32q_1^2-13m_2)\ mod \ 64)\nonumber\\
m_1&=& 64 k
\end{eqnarray}

\noindent for  leptonic Ramond representations,

\begin{eqnarray}
M_1^2&=& \frac{k}{p}\nonumber\\
M_2^2&=& \frac{X}{16p}\nonumber\\
X&=& (q_1^2-12km_1 +2(2p_2-q_2))\ mod \ 16\nonumber\\
2p_1-q_1&=& 8k
\end{eqnarray}

\noindent For N-S  $I=0$ representation  and

\begin{eqnarray}
M_1^2&=& \frac{k}{p} \nonumber\\
M_2^2&=& \frac{X}{16p}\nonumber\\
X&=& (q_1^2-12km_1 +2(m_2-2p_2+q_2))\ mod \ 16\nonumber\\
2p_1-q_1-m_1&=& 8k
\end{eqnarray}

\noindent for N-S $I=1$ representation.

\vm

 It is relatively easy to fit masses of known elementary particle and
hadrons  by choosing suitably the condensation level (presumably $p$  is
near prime power of $2$). \\ a) The dominant contribution to mass formula
comes from $ M_1^2$ and the integer appearing in this term is uniquely
fixed by the fitted mass.
 For bosons $u(1)$ charge gives no contribution to this terms and for
lowest lying mesons and gauge bosons this contribution turns out to be
vanishing.  For higher excitations of mesons on Regge trajectories this
contribution must be nonvanishing and the integer $k$  be expressed in
terms of particle spin. In lowest order the lowest order contribution is
linear in spin and one obtains the mass formula of the old fashioned
string model and prediction for the hadronic string tension. \\ b) The
second order term has natural relative accuracy $1/64M^2$ if the integers
appearing in it are {\bf  small \/} and with a
 proper choice of the parameters rather accurate fit is obtained:  the
accuracy is the better the larger the fitted mass is.   If one poses no
constraints on the values of integers one can in fact select the
parameters so that  $O(1/p^2)$ accuracy is achieved! The point is that
p-adic numbers modulo $p$ form a finite field and one can always choose
the integers appearing in the mass formula so that   the number $K$ in
$M_2^2= Kp^2$ is any p-adic integer in the range $1,p-1$:  this meas that
second order can have any value in the range  $(0,1/p)$ with accuracy
$O(1/p^2)$.  This result implies that the concept of simplicity  is
doomed to be somewhat subjective.

\subsection{ Are the real counterparts of $P,Q,m$ invariant in
secondary condensation?}

The mass formulas provide a test for the hypothesis that the real
counterparts
 of $P,Q,m$ remain invariant in topological condensation and condensation
occurs only provided the mass decreases in condensation.  The lowest order
contribution to mass squared is linear in the quantum numbers
$p_1,q_1,m_1$ and does not change but already the second order
contribution changes since the algebra of p-adic numbers is not isomorphic
with the algebra of real numbers.

\vm

Consider condensation of level $p_a<p_b$ on level $p_b$.  The first
constraint
 comes from the requirement that first order contribution to the real
counterpart of $P,Q,m$ remains invariant in order $O(1/p_2)$. Writing the
first order contribution in the general form $n/p$ the invariance gives

\begin{eqnarray}
n_b&=& \frac{p_b}{p_a}n_a+ O(\frac{1}{p_2})
\end{eqnarray}

\noindent Since $n_2$ must satisfy a condition of form $n_a \ mod \ k=0 $,
 $k=8$ or $k=64$ in order to avoid Planck mass the only manner to satisfy
the constraint   is to assume that the ratio ${p_b}{p_a}$ is integer in
accuracy of $O(1/p_1)$.

\begin{eqnarray}
\frac{p_2}{p_1}&=& r_1 +O(\frac{1}{p_1})
\end{eqnarray}

\noindent  This condition together with the condition $p_i \ mod  \ 64 =63$
 restricts the set of allowed primes considerably since condensation to
wrong levels implies Planck mass.

\vm

In second order approximation the invariance for the real counter parts to
 second   order terms $n_2\in \{m_2,p_2,q_2\}$ implies the transformation
property

\begin{eqnarray}
n^b_2&=& r_1^2n_2^a+  r_2n^a_1+O(\frac{1}{p^2_b})\nonumber\\
r_2&=& int(p_b(\frac{p_b}{p_a}-r_1))
\end{eqnarray}

\noindent Second term corrects the error associated with lowest order
term.  In principle these formulas allow to estimate the change in mass
squared to order.

\vm

The contribution of $O(p)$ terms to mass formula are linear in
 quantum numbers $p_i,q_i,m_i,q_K$ and does not change in condensation.
 An important factor affecting the situation is the appearence of $k\in
\{1/8,1/64\}$ factor,  assumed to be essentially invariant in condensation
 ($p_i \ mod \  64=63$).
 The coefficient $X$  of the second order contribution contains two parts.
 The part  $X_1$ quadratic in quantum numbers $p_1,q_1,m_1$   gets
multiplied with $r_1^2$.  The part  $X_2$ linear in $p_2,q_2,m_2$ is mixed
with the first order term of similar form  and therefore  transforms  as $
X_2 \rightarrow r_1^2X_2+ r_2s $,  where  $s$ is the coefficient of the
first order term.
  The inhomogeneous term in fact cancels the change of first order term if
the condition  $ s \ mod  \ k=0$ is preserved on condensation.
Neglecting the complication coming from the modulo  conditions this gives
for the change of  the  mass the following formula

\begin{eqnarray}
M_1^2 &\rightarrow& M_1^2\nonumber\\
M_2^2 &\rightarrow &M_2^2 r_1\nonumber\\
r_1&=& int(p_b/p_a)
\end{eqnarray}

\noindent Mass increases in condensation!

\vm

The picture is however complicated by the fact that the integer valued
coefficient $X$  is defined only  modulo $k$ equal $16,32$ or $64$ for
small values of integers appearing in it.  Furthermore,  $X$ contains
terms,  which are determined only modulo $2,4,16,32,k$ and can have either
positive or negative sign so that if $r_1^2$ is  multiple
 of $2,4,16,...$  the corresponding terms disappear from the mass formula
and mass either increases of decreases depending on the sign of the
disappearing term.  If the contribution becomes p-adically negative it has
essentially the maximal value $1/p_b$.    This phenomenon occurs for
primes near powers of two and satisfying the condition $p \ mod \ 64 =63$.
More generally,    primes near  powers of $2$ satisfy this condition. In
particular,  for primes near prime powers of  two $r_1^2$ is some power of
$2$ and the sign of the condensation energy for to nearby prime powers of
$2$ is sensitive to the values of integers $p_1,q_1,m_1$ and this
sensitivity.

\vm

   An especially interesting situation occurs if  $r_1^2 \ mod \ k=0$.  In
 this case the  contribution of the second order term seems to disappear
totally from the mass formula!      The condition is satisfied for the
condensation of Mersenne prime $p_a=2^{m_a}-1$  on Mersenne prime
$p_b=2^{m_b}-1$    since   one has  $r_1=2^{m_b-m_a}$ when the ratio of
Mersenne primes is sufficiently large.   The situation is essentially same
 for all primes  $p \ mod \ 64 =63$ near powers of two.
   If the interpretation is correct it would mean that second order
contribution
 to mass  disappears in few condensations if one assumes that physically
interesting
 primes correspond to primes near prime powers of two and the first order
term in mass is the only  stable contribution to particle mass. If first
order term to mass squared vanishes particle can lose  its entire  mass via
condensation!          Pions and intermediate gauge bosons indeed turn out
to be examples of particles for which
 first order contribution to mass vanishes.  Perhaps the unstability
against decay makes the
 loss of mass via secondary  condensation
 in practice impossible for these particles.   An interesting prediction
of the scenario is that  the value of renormalized  mass  after several
secondary condensations approaches always to the value obtained by
dropping  the second  order contribution from the mass after primary
condensation.  This prediction is
 true for any  model of condensation  based on assumption that first order
contribution to mass squared  remains invariant in condensation.

\subsection{Why primes near prime powers of two are favoured as primary
 condensation levels?}

 The difficult questions  considered already earlier  are: \\ a) Why primes
near prime powers of $2$ seem to be favoured as primary condensation
levels? \\ b) What determines the primary condensation level $p_{cr}$. \\
  The hypothesis about the invariance for the  real counterparts
 of $P,Q,m$  are invariant in condenstaion   allows to reconsider these
problems  on more quantitative level than  previously.

\vm

Let us reconsider first the problem a).
 Assume that the primary condensation level of the  particle is not unique
so  that also some primes $p<p_{cr}$ are possible as  primary condensation
levels.    Assume
 however that
    particle  mass is  more or less  independent on condensation level.
Given  prime $p_{cr} \simeq 2^k$ with
 temperature $T_{p_{cr}}=1$,   then for all primes $p_1\simeq 2^{k_1}$,
with $k_1$ a
 factor of $k$,  the thermal masses are in good approximation same if the
temperature $T_{p_1}$   is choosen to be $1/T_{p_1}= k/k_1 \in Z$.    For
$p_1=2$ this requirement can be satisfied  always so that 2-adic
condensation level is in some sense fundamental.  For $p=2^{k}$, $k $
prime,
   only $p_1=2$ satisfies the constraint so that mass depends strongly on
condensation level  irrespectively how one chooses the temperature:
amusingly  these are just the physically interesting condensation levels!
The condition implies  that for $p>p_{cr}$ the inverse temperature had to
be fractional number and this is not possible since fractional powers of
$p$ does not exist p-adically. Therefore there is maximal value $p=p_{cr}$
of $p$  for which thermal description  makes sense and this prime
characterizes particle's  mass scale.

 \vm

 At the level of moduli space the choise  $T=1/k$  means that the allowed
points  of p-adic moduli space correspond to $q_{ij}= p^{k_{ij}}$,
$k_{ij} \ mod \ k=0$: the restriction is modular invariant.     The
assumption implies that 2-adic level is the fundamental level and the
value of the  inverse temperature at 2-adic level determines the mass of
the particle.  For  primes near prime powers of two  the 2-adic inverse
temperatures are primes:

\begin{eqnarray}
\frac{1}{T_2}&=&  prime
\end{eqnarray}

\noindent This explanation for favoured condensate levels   looks very nice
mathematically.  What remains to be
 understood is how the prime $k$ depends on the quantum numbers of
particle:  why $k=127$ electron, $107$ for u and d quarks,  $89$ for
intermediate bosons,... ?

\vm

 It is useful to recall the second argument related to the real
counterparts
 of the coupling constants defined  by canonical identification as $g^2_R=
(g^2p)_R$. Mersenne
 primes have the special property that if p-adic  $g^2$ is finite
superposition of negative
 powers of two then its real counterpart is numerically equal to p-adic
counterpart and real and
 p-adic theories do not differ drastically. It remains to be seen whether
there exists also other
 primes near prime powers of two with the  same property.   For arbitrary
primes $g^2_R$ can differ  widely from $g^2$ numerically if $g^2$ is
rational number, which does not depend on $p$. If
 this is indeed the case then Mersenne primes and primes near prime powers
of two might be a  result  of a 'natural  selection'.

\vm

  Consider next a possible  mechanism determining the value of $p_{cr}$.
The
 simplest possibility one can imagine is that for $p_{cr}$ the secondary
 condensation on  nearby condensate level $p$ not much larger than
$p_{cr}$ is energetically more favoured than primary condensation  on
level $p$ so that  particle is replaced with new structure:  particle
condensed on small piece of $p_{cr}$ condensate level.  This leads to
essentially constant mass squared $\propto 1/p_{cr}$ independent of
condensate level instead of thermal mass squared proportional to $1/p$.
\\ a)  Consider two primes $p_a<p_b$, which are near to each other:
$p_b/p_a<2$.  Both $p_a$ and $p_b$ can serve as primary condensate level
for particle and particle masses at these level are related by scaling
$m_b^2/m_a^2= p_a/p_b$ if the condition $p \ mod \ 64=63$ is satisfied.
\\ b)  Particle at $p_a$ level can also suffer secondary condensation to
the level $p_b$.   In this case the changes of the integers $p_1,q_1,..$
are  small.   Writing the first order term in form $ap/k$  the condition
that particle doesn't get Planck mass in condensation   reads $
int((p_b/p_a)a \  mod \ k=0$  so that $a$ changes by a multiple of  $k$.\\
 c) This would be
  guaranteed if the relative size of change of $p$ is of order $O(1/p)$ so
that $q_1,p_1,m_1$ remain invariant and only $m_2,p_2,q_2$ change, in
which case the particle mass changes,  if  the coefficient $X_2$ linear in
these integers becomes proportional to $k= 64$, $16$ or $32$ appearing in
the denominator of this term so  that $X_2$ contribution from mass
disappears. This is however impossible:  $r_2=int (p_b -p_a)(p_b/p_a))
\propto 64$ in this case so that $X_2$ remains unchanged.  \\ d) The only
only possibility is that the change of $p$ is so large that $p_1$  or
$q_1$ and therefore $X_1$ can  change.  The conditions for this is are $
int(n(p_b/p_a)-n\equiv d \ge 1$ for  $n=p_1$ or $n=q_1$.  The
 change in $M^2$   caused by secondary condensation on one hand and by
the change of primary condensation level on the other hand   satisfies the
upper bound

\begin{eqnarray}
\vert \Delta M_2^2(sec)\vert&\leq &M_2^2=\frac{X_2}{p_{cr}}\nonumber\\
\vert \Delta M^2(prim) \vert &<& \frac{d}{n+d}M^2
\end{eqnarray}

\noindent Therefore under the following necessary  conditions

\begin{eqnarray}
\Delta M_2^2(sec)&<&0\nonumber\\
\frac{d}{n+d}&<&\frac{M_2^2}{M^2}
\end{eqnarray}

\noindent  energetics can favour secondary condensation.  If $M_2^2$
and/or $n$
 is small it  is quite possible that secondary condensation is not
possible at all and particle remains massless (or possesses only thermal
mass).  \\ e)  $p_{cr}$ might correspond to the smallest  prime for which
this kind of process is possible and resulting particle is long lived
enough.   This mechanism
 implies sensitive dependence  of $p_{cr}$  on the integers $p_1,q_1,....$
and therefore  on particle's quantum numbers.     If mechanism is really
correct it predicts that for stable elementary  particles the value of the
parameter $X$ should be as small as possible.

\vm

This kind of mechanism could be at work in cosmological context:  the
cooling of the  Universe  corresponds to the gradual increase of the
typical
 $p \propto \sqrt{1/T}$ associated with background and massive particles
would separate from thermal equilibrium as 'hot spots' and remain in their
own  internal temperature in the proposed manner.

\vm

This mechanism implies that particles would correspond to double sheeted
structures with $p_a$ and $p_b$  near each other.   The calculation of
 particle masses shows that  the idea about two p-adic levels very near
each other seems to be correct.
  For example,
 the condensation levels with  $u,d,s$ quarks is $k=107$  and same as for
hadrons.  An amusing coincidence is that Connes and Lott  \cite{Connes}
have proposed the description of Higgs mechanism based on noncommutative
geometry and one consequence of mechanism is double sheeted structure of
space time!   Of course,  the presence of many sheeted structures is  basic
 characteristic of the  topological condensate.

\section{p-Adic description of modular degrees of freedom}

 The success of the mass calculations give convincing support for
generation-genus correspondence.  The basic physical picture is
following.  \\ a)  Mass squared is dominated by boundary  contribution,
which is sum of cm and modular contributions: $M^2=M^2(cm)+M^2(mod)$.
 Here  'cm' refers to  the center of mass of the boundary component.
   Modular contribution can be assumed to depend on the genus of the
boundary component only. \\ b) Modular contribution to mass squared can be
estimated apart from overall proportionality constant.  Elementary
particle vacuum functionals are proportional to product of all even theta
functions and their conjugates the number of even theta functions  and
their conjugattes being $2N(g)= 2^{g}(2^g+1)$.   Also the thermal
partition function must also be proportional to $2N(g)$:th  power of some
elementary partition function.  This implies  that  thermal/ quantum
expectation $M^2(mod)$ must be proportional to $2N(g)$. Since single
handle behaves effectively as particle the contribution must be
proportional to genus $g$ also. The surprising success of the resulting
mass formula shows that the argument is correct.

\vm

The  challenge is to construct theoretical framework reproducing
 the modular  contribution to mass squared.  There are two alternative
manners to understand modular contribution. \\ a)  Modular contribution is
regarded as quantum mechanical expectation value of mass squared operator
for elementary particle vacuum functional. Quantum treatment would be very
straightforward in principle: generalize the concept of  moduli space and
theta function to  p-adic context and find an acceptable definition for
mass squared operator.  \\ b)  Modular contribution is calculated using
thermodynamical treatment.

\vm

 Thermodynamical treatment might go along following lines. \\
 a)  The minimal requirement is that the thermal expectation value  for
mass squared for modular  invariant partition function is well defined.
Probably the expectation
 value should be defined as logarithmic derivative with respect to
temperature type parameter appearing in the partition function. \\ b)  The
structure of elementary particle vacuum functionals suggest that partition
function must be expressible as product of $2N(g)$ elementary partition
functions. Since cm part contains all information about standard quantum
numbers it seems useless to assume any electroweak structure for $Z(g)$.
The fact that theta functions are expressible in terms of exponentials
suggests that  $Z$ is more or less
 equal to square of the elementary particle vacuum functional so that
quantal and thermal approach seem to lead to same end result.  This ansatz
guarantees automatically modular invariance provided one can somehow
modify the concept of moduli so that theta function or at least modular
invariant combinations of them become well defined concepts.  \\ c)
Modular contribution to mass squared ought to be
 small in Planck mass scale and this is possible provided the concept of
low
 temperature phase makes  sense for modular invariant partition
functions.  This constraint is strong.   The first alternative which comes
into mind is that the concept  of low temperature phase generalizes and
leads to the  quantization of allowed values of modular parameters
(analogous to temperature parameters) so that moduli space is effectively
discretized.  This picture seems to work and implies
 also that theta function becomes a well defined concept.

\vm

The realization of either of these  scenarios necessitates the
generalization
 of ordinary Riemannian geometry to p-adic
 context in some sense.
  This would mean that boundary
 component is described as p-adic complex manifold, whose coordinate space
can be imbedded to 4-dimensional square root allowing extension of p-adic
numbers.  The naive guess is that all formulas of Riemann geometry
generalize as such:  in particular the moduli space of p-adic Riemann
surfaces is obtained simply by replacing its  complex coordinates with
their p-adic counterparts.  It will turn out  that p-adic existence
requirements make the generalization far less trivial. It however seems
that the concept of low  temperature phase allows modular invariant
generalization.

\subsection{p-Adic moduli space}

It is not at all obvious that the concept of moduli space generalizes to
 p-adic context in any sensible manner.  Whatever the generalization is it
should allow the p-adic version of  elementary particle vacuum
functionals. A further constraint comes from the requirement that low
temperature phase is defined in some sense, which presumably means
discretization of moduli space.

\subsubsection{$\Theta$ function for torus}

It is instructive to first study the problem for torus first.  The
ordinary moduli space of torus  is parametrized by single complex
 number $\tau$.  The points related by $SL(2,Z)$ are equivalent, which
means that the transformation $\tau \rightarrow ( A\tau +B)/(C\tau+D)$
produces a point equivalent with $\tau$. These transformations are
generated by the shift $\tau \rightarrow \tau +1$ and $\tau \rightarrow
-1/\tau$. One can choose the fundamental domain of moduli space to be the
intersection of the slice
 $Re (\tau) \in [1/2,-1/2]$ with the exterior of unit circle $\vert \tau
\vert =1$.     The idea is to start directly from physics and  to look
whether one might some define p-adic version of elementary particle vacuum
functionals  in  the p-adic counter part of this set  or in some modular
invariant  subset
 of this set.

\vm

Elementary particle vacuum functionals are expressible in terms of theta
 functions and a good  guess is that  this holds true for  partition
function, too.  The general expression for theta function reads as

\begin{eqnarray}
\Theta [a,b](\Omega)&=& \sum_n exp(i\pi(n+a)\cdot\tau \cdot (n+a) +
 2i\pi(n+a)\cdot b)
\end{eqnarray}

\noindent  $a$ and $b$ are half odd integers for torus.  The  obvious
problem is  that $\pi$ does not exists p-adically  and one should somehow
make sense of the theta function perhaps somehow modifying the concept of
modular parameter.  Also the concept of low temperature phase ought to
make sense.  The idea is to regard $q\equiv exp(i\pi\tau)$ as a basic
coordinate in the fundamental domain instead of $\tau$ so that one gets
rid of the problems related to the nonexistence of $\pi$.

\vm

Consider first the contribution of the real part of $\tau $ to exponent
appearing in $q$  assuming  that $ Re( \tau)$ belongs to the fundamental
domain.
 One can always right $Re (\tau)$ in the form  $(m/n)p^{-k}$, $n>1$, $k\ge
0$   in the fundamental domain and it corresponds  to p-adic number with
norm not smaller than one so that  the exponent does not converge as power
series and the only possible interpretation for $q$  is
 as a complex root of unity that is
 as  a p-adic   complex number $x+iy$ satisfying the condition $x^2+y^2=1$,
  from which one gets $y= \sqrt{1-x^2}$. Series converges if the p-dic
norm  $x$ is  smaller than one so that one has $x \ mod \  p=0$.

\vm

 An additional  discrete set of phases is  obtained  by requiring that $y$
and $x$ are also finite with respect to real topology. With this
assumption  phases of  form $ (k/l) +i(m/n)$ satisfying
$(k/l)^2+(m/n)^2=1$  are allowed for $q$.  The condition is equivalent
with the age old problem  considered already by Babylonian mathematicians
of finding solutions of   $k^2+l^2=n^2$ that is finding integer sided
rectangular triangles (Pythagorean triangles). Solutions can be regarded
as complex numbers of $k+il$ and form monod with respect to
multiplication.   The condition $Re(\tau) \in [-1/2,1/2]$ for fundamental
domain of torus gives $ -\pi \leq\Phi\leq \pi$ so that all possible
triangles are allowed in  the fundamental domain.   The action of the
modular
 transformations $\tau \rightarrow A\tau D^{-1}+BD^{-1}$ on these phase
factors  is trivial for torus.  For higher genera the action is  just
$exp(i\Phi)\rightarrow  exp(in\Phi)$,  which  belongs to the allowed set
since the allowed phases form a monod with respect to multiplication.   By
a little calculation one verifies  that the explicit form for the allowed
$(k,l)$ and $(l,k)$ pairs is given by

\begin{eqnarray}
k &=&2rs\nonumber\\
l&=& r^2-s^2\nonumber\\
n&=& r^2+ s^2
\end{eqnarray}

\noindent where $r$ and $s$ are relatively prime integers,  not both odd.
 Note that $(l,k)$ is also an allowed solution. An important point to be
noticed is that the p-adic norm of $exp(i\Phi)$ is not larger than one for
physically interesting primes satisfying $p \ mod \ 4 =3 $ since  $ n \
mod \ 4=1$ holds true as a simple calculation shows.

\vm

A possible source  of problems is the appearence of  terms  $exp(i\Phi/4)$
in the definition of theta function,  when $a$ is half  odd  integer.  The
phase $exp(i\Phi/4)$  however separates into a multiplicative factor and
since elementary particle vacuum functional is proportional to the product
of thetas and their conjugates these ill defined phases cancel each
other.     Note that the the phase factor $exp(i2i\pi(n+a)\cdot b)$ is for
even theta functions appearing in the definition of elementary particle
vacuum functional always equal to $\pm 1$ and therefore well defined.

\vm

An additional constraint to the allowed  phase factors  comes from the
requirent
 that the integral  (or rather discrete sum) of the square of elementary
particle vacuum functional over moduli space converges.  The problem is
that infinite number of phases contribute to the sum and it is not clear
whether the sum converges.  The problem is that the p-adic norm of phase
factor defined by Pythagorean triangle has p-adic norm equal to one.

\vm

A certainly working scenario is based on the assumption that the phase of
$q$ is completely trivial.  This  means that  only diagonal metrics $ds^2=
dx^2+ Im (\tau)^2dy^2)$ are allowed. The restriction to trivial phase
factors respects modular invariance.

  \subsubsection{Imaginary part of the period matrix}

Consider next the contribution $exp(-2\pi Im(\tau)$ of the imaginary part
 of $\tau$ to $q$.  The only sensible manner to define this quantity is to
require that the modulus of  $q$ rather than $Im(tau)$ is the  fundamental
quantity just like the phase of $q$ rather than $Re(\tau)$ is taken as
fundamental quantity.  The constraint that $ Im (\tau)$  corresponds  in
ordinary complex case the constraint $\vert q \vert < exp( -2 \pi \sqrt{(1-
(m/(np^k))^2})$ for $Im(\tau)>0$ (standard convention) or
 $\vert q \vert >exp( 2 \pi \sqrt{(1- (m/(np^k))^2)}$ if one takes $Im
(\tau)$ to be negative.  The subset of allowed p-adic values of $q$ should
satisfy this constraint in some sense: as such the constraint does not
make sense p-adically.   The
 $\pm$ sign is present since one can choose between two alternatives
$Im(\tau)<0$ or $Im(\tau)>0$ in complex case.

\vm

The requirement that low temperature phase is in question gives an
additional
 constraint. Temperature quantization means in ordinary thermodynamic
context that the exponent $exp(1/T)$  reduces to integer power of $p$.
The essential requirement is that the exponents $q^n$ are of order
$O(p^{kn})$, $k>0$,  p-adically so that the power expansion of eta
function  with respect to $q$ converges extremely rapidly. This is
achieved by requiring

\begin{eqnarray}
N_p(mod(q))&<&1
\end{eqnarray}

\noindent This condition means that one can write $mod(q)= p^k (m/n)$,
$k>0$,
 where $(m/n)$  is p-adic rational number with unit p-adic norm.
Condition is  modular invariant since all modular transformations lead
outside the fundamental domain.  It turns out that a stronger condition
$q=p^k$ is needed in order to obtained well defined partition functions.

\subsubsection{Discretization of moduli space}

Restrictions on the values of the allowed modular parameters are probably
 necessary since  the integration over p-adic  moduli parameters is not
obviously well defined concept.  \\
 a) Discretization of moduli  of $q$

\begin{eqnarray}
mod(q)&=& p^k, k>0
\end{eqnarray}

\noindent replaces integrations over $mod (q)$ with summations and leads
directly to the
 thermodynamic picture with the difference that all temperatures $T
\propto 1/k$ are allowed.\\ b) The assumption $q=p^k$ so that  the phase
of $q$ is completely trivial leads to mathematically well defined
expressions for partition functions and to strict thermal interpretation
of the partition function  in the sense that  temperature parameters are
real.  If phases defined by Pythagorean triangles are
 allowed the integration gives for for each power of $p$ an infinite sum
over phases, which have p-adic norm equal to one and it is not clear how
the convergence of the sum could be achieved.

\subsubsection{Higher genera }

The generalization of the low temperature phase moduli space to the case
of  higher genera seems rather obvious. The rule is  simple:  $q_{ij}=
exp(i2\pi \tau_{ij}) \rightarrow p^{k_{ij}}exp(\Phi_{ij})$, $k_{ij}>0$,
where either of
 two possible scenarios for the phases is assumed.   This definition is
 modular invariant if all modular transformations not leading  out of this
set respect this condition.   The transformations permuting different
nonintersecting homology cycles permute the rows of Teichmuller matrix and
are not problematic.  Also the  transformations $\Omega \rightarrow A\Omega
D^{-1}$  leave this set invariant.      The guess is that these are the
only modular transformations not leading out of the discrete moduli space.

\vm

Since $Im (\Omega)$ and
 $Re(\Omega)$ transform independently in modular tranformations  in
question the determinant $det(Im(\Omega))$ or equivalently the integer
valued determinant

\begin{eqnarray}
D_1&\equiv& det(k_{ij})
\end{eqnarray}

\noindent  is  p-adically well defined  modular invariant.   This implies
that one can perform gauge fixing in moduli space  in order to avoid
multiple summations in the construction of partition function. All what is
needed is to pick up one representative  $(k_{ij},exp(\Phi_{ij}))$ from
each orbit of the modular group.  The orbits are labeled by the value of
determinant of $D_1$ and all allowed values of the phase  factor
$exp(\Phi_{ij})$.  Later considerations show that one must assume

\begin{eqnarray}
D_1&\ge& 0
\end{eqnarray}

\noindent  for the allowed modular parameters in order to achieve
convergence
 of the partition function.

\subsection{Elementary particle vacuum
functionals in  p-adic context}

The general definition of theta function $\Theta [a,b](\Omega)$ for  genus
$g$, where $a$ and $b$ are vectors having $g$ components with values in
$\{0,1/2\}$ reads as

\begin{eqnarray}
\Theta [a,b](\Omega)&=& \sum_n exp(i\pi(n+a)\cdot\Omega \cdot (n+a) +
 2i\pi(n+a)\cdot b)
\end{eqnarray}

\noindent For discrete moduli space $exp(i\pi\Omega_{ij})=p^{k_{ij}}$ this
quantity is indeed well defined.  The only source of troubles is the term
 $exp(2 i\pi (n+a)\cdot b)$.  The quantity $exp(i2\pi n\cdot b)$  can be
defined as
 $\pm 1$ depending on whether the inner product $n\cdot b$ is integer or
half odd integer.   For even theta functions appearing in  elementary
particle vacuum functionals the inner product $a\cdot b$ is half odd
integer so that one obtains $-1$ from this term. This means that even
theta functions are real quantities in discrete moduli space  and can be
written in the form

\begin{eqnarray}
\Theta [a,b](\Omega)&=& \sum_n p^{(n+a)^ik_{ij} (n+a)^j }\epsilon (n)
\nonumber\\
\epsilon (n)&=& -exp(2i\pi n\cdot b) (=\pm 1)
\end{eqnarray}

\noindent  As far as practical calculations are considered only few terms
are important due to the extremely rapid convergence for large values of
$p$.

\vm

The  modular invariant vacuum functional can be defined  just as in
\cite{TGD}

\begin{eqnarray}
\Omega (vac)&=&
\frac{\prod_{a,b}\Theta [a,b](\Omega) \bar{\Theta} [a,b](\Omega)}{(\sum \Theta
[a,b](\Omega)\Theta \bar{\Theta} [a,b](\Omega))^{N(g)}} \end{eqnarray}

\noindent The product in the numerator is over all even theta functions.
The nice feature of the definition is that vacuum functionals vanish for
moduli possessing conformal symmetries: this is essential for the argument
about 3 light fermions generations.

\subsection{Definition of mass squared expectation value}

  The
general form of the theta function suggests a natural definition
 of the mass squared expectation value.  Theta function is analogous to a
discrete version of functional integral.  Functional integration
corresponding to summation over  over multi-integers $n$ and the
counterpart of action is the quadratic form $S(a,n,k_{ij})=(n+a)^i
k_{ij}(n+a)^j$ so that 'free'  theory is in question. The action of the
mass squared operator on $\Theta [a,b]$ should correspond to functional
expectation value of the action $2S(a,n,k_{ij})$ over $n$ (note the
appearence of factor $2$, which turns out to be necessary)

\begin{eqnarray}
M^2\circ \Theta [a,b](k_{ij})&\equiv& \sum_n  2S(n,a,k_{ij})
 p^{S(n,a,k_{ij})
}\epsilon (n) \nonumber\\
S(n,a)&=&(n+a)^ik_{ij} (n+a)^j
\end{eqnarray}

\noindent  In the expectation value of mass squared operator for vacuum
functional $M^2$ operator should and like differential operator on   the
vacuum functional/partition function. This can be formally achieved by
introducing formal temperature parameter by scaling the 'action' $S$
appearing in the definition of theta function by temperature and defining
the action of mass square operator as a logarithmic derivative with
respect to $T$

\begin{eqnarray}
\Theta [a,b](k_{ij},T)&\equiv& \Theta [a,b] (\frac{k_{ij}}{T})\nonumber\\
M^{2} \circ F &= & -2\frac{T}{dT} F_{\vert T=1}
\end{eqnarray}

\noindent  This definition is in accordance with the idea that mass squared
operator,  or essentially Virasoro generator $L_0$,  measures the action
of an infinitesimal scaling: now the scaling acts on modular parameters
rather than complex coordinates. Since the contribution of modular degrees
of freedom changes only the vacuum weight of Super Virasoro representation
there is no necessity to define the action of $L_n$, $n\neq 0$ on vacuum
functional.

\vm

 Quantum mechanical and thermal   mass expectation
 values are identical provided one identifies thermal partition function
$Z$  in suitable manner

\begin{eqnarray}
\langle M^2\rangle_{qu} &=& \frac{\int dV \Omega  M^2 \circ \Omega}{\int dV
\Omega^2}= -2\frac{T}{dT} Z_{\vert T=1}\nonumber\\
Z(T)&\equiv& \int dV \Omega^2 (k_{ij},T)
\end{eqnarray}

\noindent  Integration measure $dV$
 corresponds actually to the summation over the nonequivalent points of the
moduli space, one point for each value of $det (k_{ij})$.

\vm

 It remains to be shown that the  action of mass squared operator on theta
function is modular invariant.  A straightforward calculation shows that
the lowest order contribution
 to mass squared expectation value is what it is expected to be.  Consider
torus as an example.  Even  theta functions
 correspond to  characteristics  $[a,b]=[1/2,0],[0,1/2]$ and $[0,0]$ and
their  explicit expressions read as

\begin{eqnarray}
\Theta [1/2,0](p^k)&=& p^{k/4}\sum_n p^{k(n^2+\frac{n}{2})}\nonumber\\
\Theta [0,1/2]&=& p^{k/4}\sum_n p^{kn^2}(-1)^n\nonumber\\
\Theta [0,0]&=& p^{k/4}\sum_n p^{kn^2}\nonumber\\
\end{eqnarray}

\noindent  Notice that one must have $k>0$ for the allowed powers of $p$.
A  little calculation shows that leading order term in  vacuum functional
reads as

\begin{eqnarray}
\Omega_{vac}&\simeq& K(1+3p^k)\nonumber\\
K&=&27
\end{eqnarray}

\noindent  Clearly the   $p^k=p$ of moduli space gives the dominating
 contribution to mass squared expectation.  In principle there is
$O(p^{k/2})$ term present but the contributions  coming from  numerator
and denominator cancel.  The coefficient $K$ doesn't affect  the
expectation value and lowest order contribution to mass squared can be
read directly from the coefficient of $O(p)$ term and one has $M^2=
6=2N(g)g$.  Factor $2$ comes from the definition of  $M^2$ operator.

\vm

 Mass expectation is  of order $O(p)$  for $T=1$ unless the point of
moduli space giving dominating contribution  happens to be point with
conformal symmetry: in this case the contribution to mass squared operator
becomes $O(1)$ since the
 integral in denominator is of order $O(p)$ in this case.  There is
extremely rapid converge with respect to powers of $p$ and the calculation
of thermal average is a simple task.   One can even define what $T=1/n$
expectation value  means   by restricting the points of discrete moduli
space so that  $k_{ij}$ is multiple of $n$.

\vl

\begin{center}
{\bf Acknowledgements\/}
\end{center}

\vm

It would not been possible to carry out this work without the  concrete
help of
 my friends  in concrete problems of the everyday life and I want to
express my gratitude to  them.  Also I want to thank J. Arponen,  R.
Kinnunen, J.  Maalampi and P.Ker\"anen  for practical help and interesting
discussions.


\newpage


\begin{thebibliography}{99}
\bibitem[Abe {\it et al}]{Abe}
F. Abe {\it et al}(1944),  Phys. Rev. Lett. 73,1
\bibitem[Allenby and Redfern]{Number1}
R.B.J.T. Allenby and E.J. Redfern (1989), {\em Introduction to Number
Theory with Computing}, Edward Arnold
\bibitem[Borevich and Shafarevich]{padic}
Z.I. Borevich and I.R. Shafarevich (1966) ,{\em Number Theory},
Academic Press.
\bibitem[Brekke and Freund]{padrev}
L. Brekke and P.G. O. Freund (1993), {\em p-Adic Numbers in Physics},
 Phys. Rep. vol. 233, No 1
\bibitem[Connes and Lott]{Connes}
A. Connes and J.Lott(1990), Nucl. Phys. B (Proc. Suppl.) 18, 29
\bibitem[Dudley]{Number}
U. Dudley (1969), {\em Elementary Number Theory}, W.H. Freeman and
Company
\bibitem[Gelfand {\it et al\/}]{Gelfand}
Gelfand, Graev, Vilenkin (1966) {\em Generalized Functions},
 Academic Press.
\bibitem[Goddard {\it et al\/}]{Goddard}
P. Goddard,A.Kent,D.Olive (1986),Commun.Math.Phys 103,105-119
\bibitem[Hrennikov]{Padprob}
A. Yu. Hrennikov, {\em p-Adic Probability and Statistics},
 Dokl. Akad Nauk, 1992 vol 433 , No 6
\bibitem[Itzykson {\it et al\/}]{Bible}
C.Itzykson,Hubert Saleur,J-B.Zuber (Editors)(1988):{\em
Conformal Invariance
and Applications to Statistical Mechanics}, Word Scientific
\bibitem[Pitk\"anen$_a$]{TGD}
M. Pitk\"anen (1990) {\em Topological Geometrodynamics} Internal
Report HU-TFT-IR-90-4 (Helsinki University).  Summary of
  of Topological Geometrodynamics in book form. Book contains construction
  of Quantum TGD, 'classical' TGD and  applications to various branches of
physics.
\bibitem[Pitk\"anen$_b$]{padTGD}
M. Pitk\"anen (1990) {\em Topological Geometrodynamics and p-Adic Numbers}.
 The book  describes the  general views concerning application p-adic
numbers to TGD.
 The mass fit of hadrons and elementary particles described in book  is
wrong
  at the level of details but basic ideas of p-adic description of Higgs
mechanism are identified
 already  in this work.  p-Adic length scale hypothesis is applied
systematically in various  length scales and the concept of p-adic
fractals with some illustrations  is introduced.
 \bibitem[Pitk\"anen$_1$]{Heavy}
 Pitk\"anen, M. (1990) {\em Are Bound
States  of Color Excited Leptons...}  Int. J. of Th. Phys.,29,275
 \bibitem[Pitk\"anen and M\"ahonen]{Lepto}
Pitk\"anen, M. and Mah\"onen P. (1992) ,  Int. J. of Theor. Phys.
 31,229
\end{thebibliography}
\end{document}